\documentclass{aa}
\usepackage{lscape}
\usepackage{graphicx}
\usepackage{txfonts}
\usepackage{lipsum}
\usepackage{subcaption}

\usepackage{lscape}

\usepackage{placeins}
\usepackage{booktabs}
\usepackage{hyperref}

\hypersetup{
    colorlinks=true,
    linkcolor=blue,
    citecolor=blue,
    urlcolor=blue
}
\titlerunning{A study of newly discovered close binary open clusters in the Milky Way}
\authorrunning{Li \& Zhu}

\begin{document}

   \title{A study of newly discovered close binary open cluster candidates in the Milky Way from Gaia DR3}


   \author{Zhongmu Li\inst{}
        \and Zhanpeng Zhu\inst{}
        }

   \institute{Institute of Astronomy and Information, Dali University, Dali 671003, China\\
             \email{zhongmuli@126.com}
}

   \date{Received 1 April 2025 / Accepted 9 July 2025}


  \abstract
   {With the release of Gaia data, the number of known Galactic open clusters (OCs) has increased rapidly, providing an excellent opportunity to confirm more binary open clusters in the Milky Way.}
   {Using a recently released OC catalogue, we employed the photometric and astrometric data of OCs and their member stars to find close binary open clusters (CBOCs).}
   {The three dimensional spatial coordinates, proper motions, and colour-magnitude diagrams (CMDs) are used for identifying candidate CBOCs. The fundamental parameters of 26 star clusters are determined by fitting CMDs to stellar population isochrones, to check  the similarity of reddenings, ages and metallicities of the sub-clusters of candidate CBOCs. The virial equilibrium is then used to exclude fake CBOCs including unbound moving groups. To further confirm the binary nature of the CBOC candidates, we calculated their Roche radii and orbital parameters. The tidal radius and radial velocity difference are then compared to the Roche radius and orbital velocity respectively, to find out gravitationally bound pairs.}
   {We identified nine new CBOC candidates from bound candidate open clusters, seven of which are shown to be candidates for primordial binary open clusters (PBOCs). However, only the pair CWNU 1024 and OCSN 82 is identified as a gravitationally bound CBOC, when considering the uncertainties. The other eight CBOC candidates appear to be gravitationally unbound pairs, but the results depend on the methods of tidal radius determination and gravitational binding examination.}
   {}


   \maketitle

\section{Introduction}

Open clusters (OCs) result from the gravitational collapse of gas and dust in giant molecular clouds. Observational evidence suggests that a small fraction of them form groups, such as pairs, triplets, or higher multiplicity systems \citep{2016MNRAS.455.3126C}, known as primordial binary open clusters (PBOCs).
PBOCs are expected to be transient, with possible evolutionary paths including merging, tidal disruption, and separation \citep{2015MNRAS.453..106D}.
Exploring the Galactic binary open clusters provides new insights into the formation and evolution of star clusters and the Milky Way \citep{2017AA...600A.106C}.

There are some possible scenarios for the formation of binary star clusters: (i) simultaneous formation, in which both clusters form from the same molecular cloud (hence with the similar age and metallicity) and are located in close proximity \citep{1997AJ....113..249F,2004ApJ...602..730B}; (ii) sequential formation, in which stellar winds or supernova shocks generated within a cluster elicit the collapse of a nearby cloud, leading to the formation of a companion cluster \citep{1995ApJ...440..666B}.
Another possible mechanism for binary open cluster formation is tidal capture, where two clusters with different ages and metallicities form gravitationally bound pairs during a close encounter \citep{2009AA...500L..13D,2021ApJ...923...21C}.
In addition, there are optical pairs that do not physically interact but merely overlap in the line of sight.

A series of works have utilized different OC catalogues to identify binary open cluster candidates. \cite{1995AA...302...86S} employed the OC catalogue published by \cite{1995yCat.7092....0L} to identify 18 candidate binary open clusters, applying a spatial separation criterion of less than 20\,pc between the two clusters.
\cite{2009AA...500L..13D} used a volume-limited sample from the WEBDA \citep{2003AA...410..511M} and NCOVOCC \citep{2002AA...389..871D} catalogues, located at the solar circle, to identify candidate binary open clusters.
The primary criterion they employed is that the pair's physical (not projected) distance must be less than three times the average tidal radius of the clusters in the Milky Way disk (10\,pc, according to \citealt{2008gady.book.....B}).
Based on this criterion, they concluded that at least 12 percent of all OCs within the solar circle appear to be experiencing some type of interaction with another cluster.
Using the catalogue data published by \cite{2020A&A...633A..99C}, \cite{2021AA...649A..54P} identified 133 clusters containing stars assigned to at least one other cluster. From these clusters they identified 60 cluster aggregates.
Using the \cite{2020AA...640A...1C} catalogue, \cite{2022AA...666A..75S} identified fourteen candidate binary open clusters by comparing the proper motions and colour-magnitude diagrams (CMDs) of different clusters, limited to a separation of 50\,pc.
\cite{2021ApJ...923...21C} used 2MASS \citep{2006AJ...131..1163} and Gaia EDR3 \citep{2021A&A...649A...1G} data to confirm a binary open cluster (NGC 1605a and NGC 1605b) formed via tidal capture, based on their close proximity and similar dynamics.
However, the result was quickly rejected, because \cite{2022RNAAS...6...58A} re-analysed the Gaia EDR3 data and found no evidence of NGC 1605 being a binary open cluster.

Although significant progress has been made in studying binary open clusters, the exact fraction of binary open clusters in the Milky Way remains unclear. Observations and simulations produce differing results. The proportion of binary open clusters in the Milky Way was estimated to be 8--12\% \citep{1995AA...302...86S,2009AA...500L..13D}, similar to that in the Magellanic Cloud \citep{1988MNRAS.230..215B,1990AA...230...11H,2002AA...391..547D}.
All of the above results are based on observed data, meaning they depend on the quality and extent of the input dataset.
However, conclusions based on numerical simulations are less Gaia-dependent or even completely independent of Gaia. In fact, Gaia is helping to confirm or reject predictions from numerical modelling. In the particular case of binary open clusters, numerical simulations \citep{2007MNRAS.374..931P,2010ApJ...719..104D,2016MNRAS.457.1339P}
suggest that star cluster binarity is a temporary, relatively
short-lived state, mainly linked to the early stages of the evolution of
star-forming regions. From there, one would expect a small but statistically significant fraction of binary (and even higher multiplicity) open clusters among the young ones, but a rather small fraction when considering the overall star cluster population.
When using a larger sample (post-Gaia), the true fraction of binary open clusters is edging closer to 1\% (instead of to the pre-Gaia 10\%), which is fully consistent with the results coming from numerical simulations (binarity for star clusters is transient).

$Gaia$-DR3 \citep{2023A&A...674A...1G} provides high-precision 5-dimensional astrometric parameters (positions, parallax, and proper motions) and three-band photometry ($G$, $G_{BP}$ and $G_{RP}$) for 1.5 billion sources.
Thanks to the vast and precise Gaia data, more OCs are being discovered (see recent works e.g. \citealt{2023AA...673A.114H,2023ApJS..265...12Q,2024RAA....24e5014L}).
In particular, \citet[hereafter HR23]{2023AA...673A.114H} provides a homogenized catalogue containing 7,167 clusters. In this work, we use the larger OC sample (HR23) to search for close binary open clusters (CBOCs).
The structure of this paper is as follows: Sect.~\ref{sec:data} presents the data used to search for new candidate CBOCs;
Sect.~\ref{sec:method} introduces the method employed in this study;
Sect.~\ref{sec:result} reports the results and analyses the properties of the newly discovered candidate CBOCs;
Sect.~\ref{sec:conclusion and discussion} summarises this work and discusses the limitations of both the methodology and data.

\section{Data} \label{sec:data}
We take the HR23 catalogue for this work, because it is one of the newest and largest OC catalogues.
A HDBSACN algorithm was used by HR23 to conduct a blind all-sky search for OCs based on 729 million sources with stellar magnitude brighter than 20 in Gaia Data Release 3 (DR3) \citep{2023AA...674A..37G}. Then the variational inference Bayesian convolutional neural network trained by unbiased and representative CMD pixel images of OC samples was used to verify the clusters identified by HDBSACN. A homogenised cluster catalogue is finally obtained by HR23, including 7167 clusters, of which 2387 are newly discovered. HR23 also provides the fundamental astronomical parameters of clusters, such as age, extinction, distance and tidal radius in the catalogue.

\section{Method} \label{sec:method}

\begin{figure}[b]
    \centering
    \includegraphics[width=0.48\textwidth]{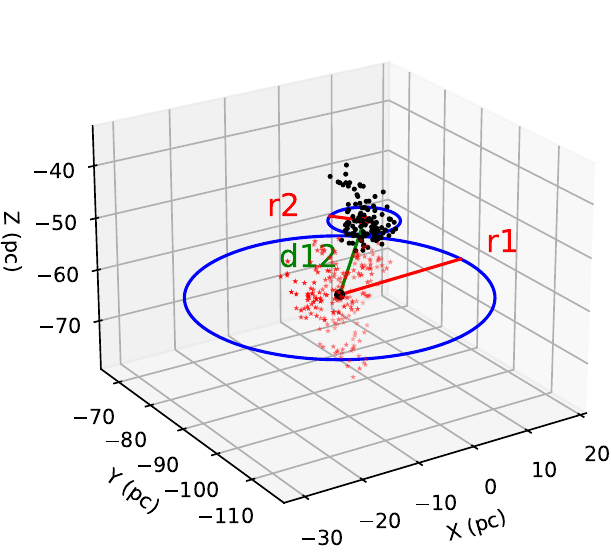}
    \caption{Close binary cluster judgement method. Two clusters with $d_{\rm 12}$ $\leq$ $r_{\rm 1}$ + $r_{\rm 2}$ are considered as a close pair. $d_{\rm 12}$ is the distance between two cluster centres. $r_{\rm 1}$ and $r_{\rm 2}$ are the tidal radii of two clusters.}
    \label{sec:fig.1}
\end{figure}

Similar to previous works, we confirm binary cluster candidates based on the similarity of two sub-clusters in proper motion and spatial distribution space, using a novel technique. In order to make it convenient to calculate the distance between two sub-clusters, we first compute the three dimensional spatial coordinates of cluster centres and member stars of clusters, from the right ascension, declination, and parallax (RA, DEC, $\omega$) data. Note that while HR23 provides the Galactocentric coordinates of cluster centres in the XYZ space, there are no similar data for their member stars. We therefore compute the X, Y, and Z values at first. We then calculate the absolute distance between the centres of two sub-clusters and check whether the cluster pair is a binary cluster candidate via the distance.
In previous works \citep{2009AA...500L..13D,2017AA...600A.106C}, the critical distance between the two sub-clusters of a binary cluster candidate was taken as different values from 30 to 100\,pc. A larger critical distance will help to get more preliminary binary cluster candidates. However, there is a disadvantage for taking only such fixed critical distances, because it cannot judge whether a pair is a close one as the sizes of star clusters are usually different. We therefore use a relative distance, i.e. the sum of total tidal radii of two sub-clusters ($r_{\rm 1}$ + $r_{\rm 2}$), to find out close binary cluster candidates. If the distance between two cluster centres ($d_{\rm 12}$) is less than the sum of the tidal radii ($r_{\rm 1}$ + $r_{\rm 2}$), i.e. $d_{\rm 12}$ $\leq$ $r_{\rm 1}$ + $r_{\rm 2}$ (see Fig.~\ref{sec:fig.1}), we consider this pair of clusters to be a close binary cluster candidate. However, if the distance is too small, the two sub-clusters possibly belong to the same cluster. In this work, two sub-clusters that are closer than 0.5 times of the minimum radius of them ($d_{\rm 12} < $ 0.5~min($r_{\rm 1}$, $r_{\rm 2}$)) are regarded as the same one. This is reasonable because the intersection of the two sub-clusters is more than half the size of the smaller cluster in such case.

\begin{figure}[b]
    \centering
    \includegraphics[width=0.48\textwidth]{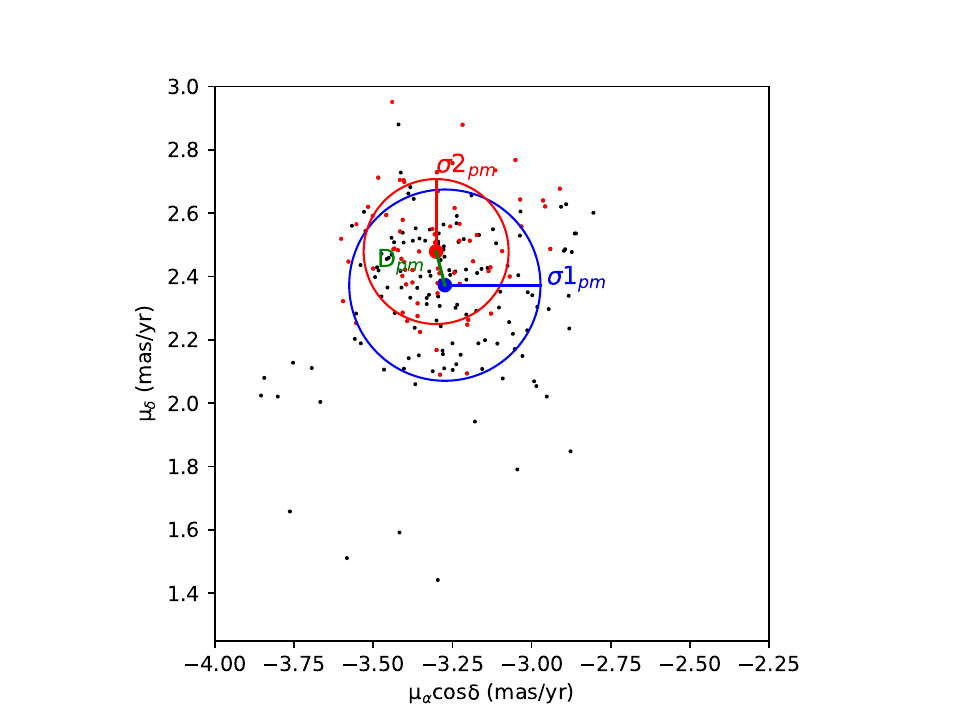}
    \caption{Proper motion similarity judgement method. Two clusters with $D_{\rm pm}$ $<$ max($\sigma$1$_{\rm pm}$,$\sigma$2$_{\rm pm}$) are thought to have common proper motion. $D_{\rm pm}$ is the distance between the proper motion centres of two clusters. $\sigma$1$_{\rm pm}$ and $\sigma$2$_{\rm pm}$ are the $\sigma$ values of two clusters.}
    \label{sec:fig.2}
\end{figure}

Besides the similar spatial position, binary cluster candidates usually have similar proper motions. \cite{2022AA...666A..75S} suggest that two sub-clusters share a common proper motion if their proper motions are consistent within 3$\sigma$ error range. In fact, such restriction is relatively lenient, which may lead to large proper motion difference. In this work, we take a new method. We compare the distance between the central proper motions ($D_{\rm pm}$) of two sub-clusters and the maximum 1$\sigma$ error in proper motion. If $D_{\rm pm}$ is less than maximum 1$\sigma$ error, i.e. $D_{\rm pm}$ $<$ max($\sigma$1$_{\rm pm}$, $\sigma$2$_{\rm pm}$), the two sub-clusters have similar proper motion (see Fig.~\ref{sec:fig.2}). This gives somewhat more strict constraint of proper motions of two sub-cluster than \cite{2022AA...666A..75S}, but it makes the results more reliable.

Through this process, we found 116 preliminary binary cluster candidates. However, we find that many clusters form groups with more than two clusters, which we refer to as cluster aggregates. In the following binary analysis we eliminated all cluster aggregates. This work investigates only the cluster pairs composed of two sub-clusters. The membership and characteristics of cluster aggregates will be examined in the future works.

In addition, we checked whether a binary cluster candidate including moving group or unbound object. We also check whether a binary cluster candidate is  gravitationally bound. The one-dimensional velocity dispersion ($\sigma_{\rm 1D}$) is compared to the velocity dispersion that is needed for virial equilibrium ($\sigma_{\rm vir}$), to check whether a cluster candidate is bound or not. The radial velocity difference ($\Delta$ $RV$) of a pair is compared to the orbital velocity ($V_{\rm orb}$) to check whether a pair is gravitationally bound, and the result is double checked by comparing the Roche and tidal radii of a cluster.

\begin{figure*}[h]
    \centering
    \includegraphics[width=\hsize]{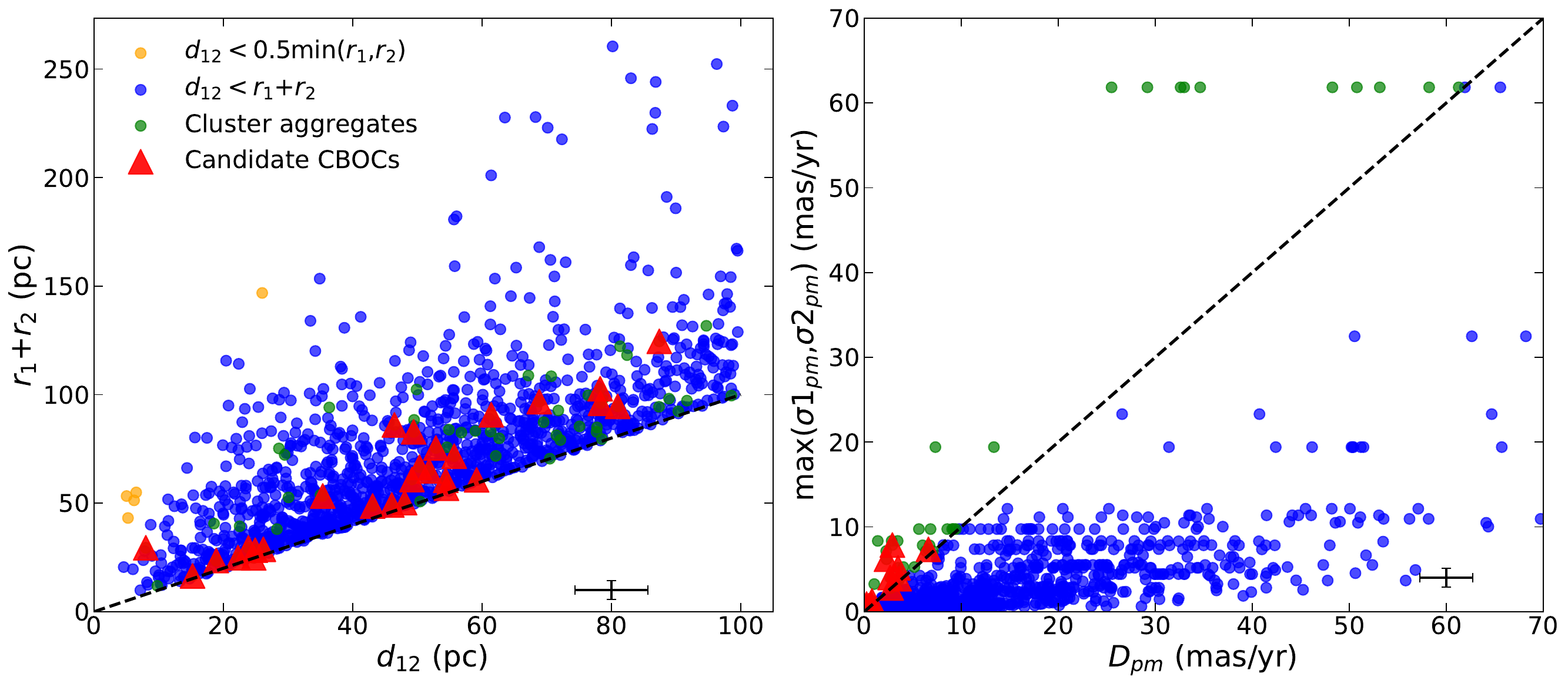}
    \caption{Classification of all cluster pairs. Orange dots indicate pairs that the member clusters may belong to the same cluster. Blue dots indicate pairs whose separations are less than the sum of the tidal radii of member clusters and whose proper motions differ clearly. Green points denote cluster aggregates. Red triangles indicate CBOC candidates.}
    \label{sec:fig.3}
\end{figure*}

\begin{figure*}[htbp]
    \centering
    \includegraphics[width=0.49\hsize]{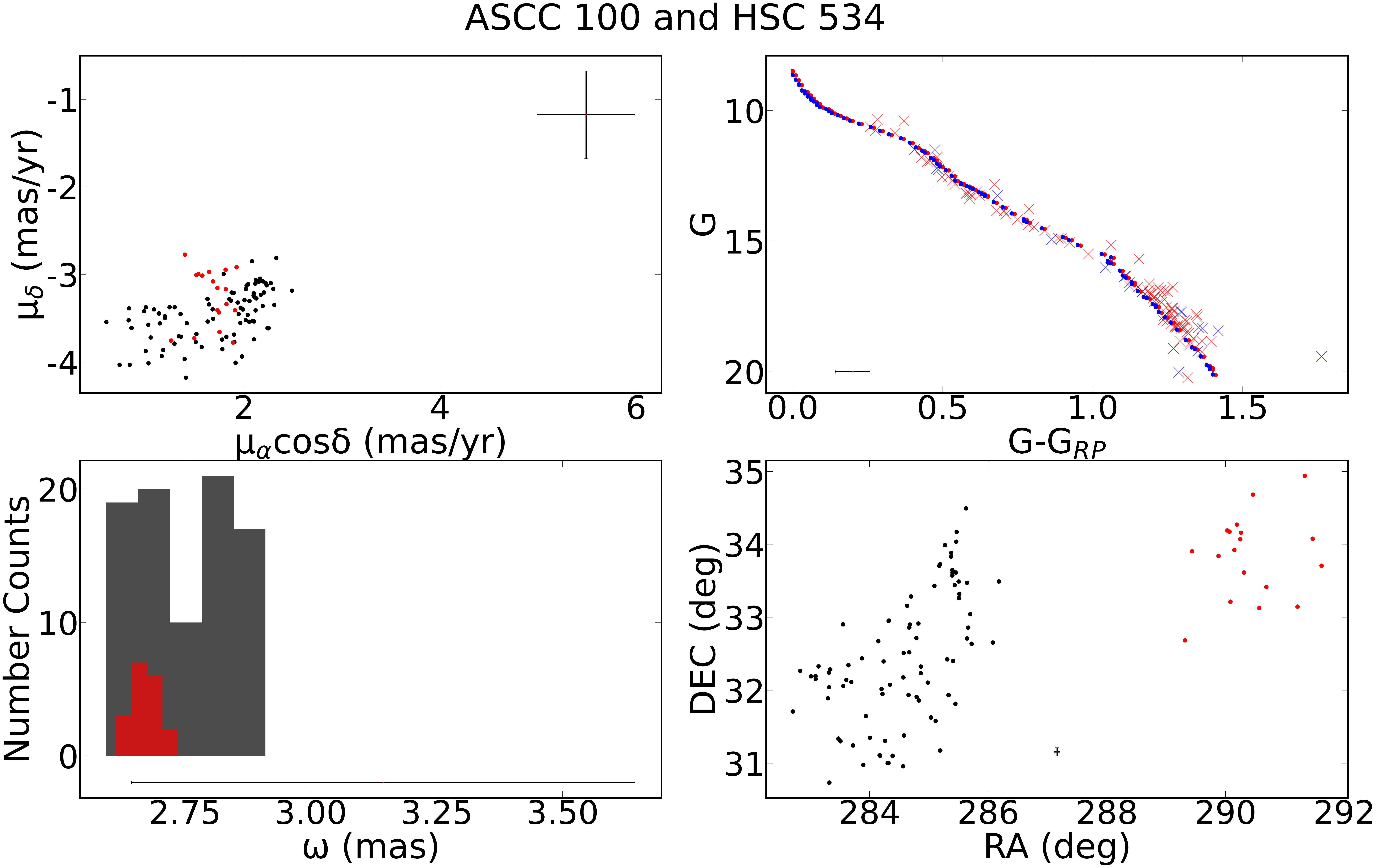}
    \includegraphics[width=0.49\hsize]{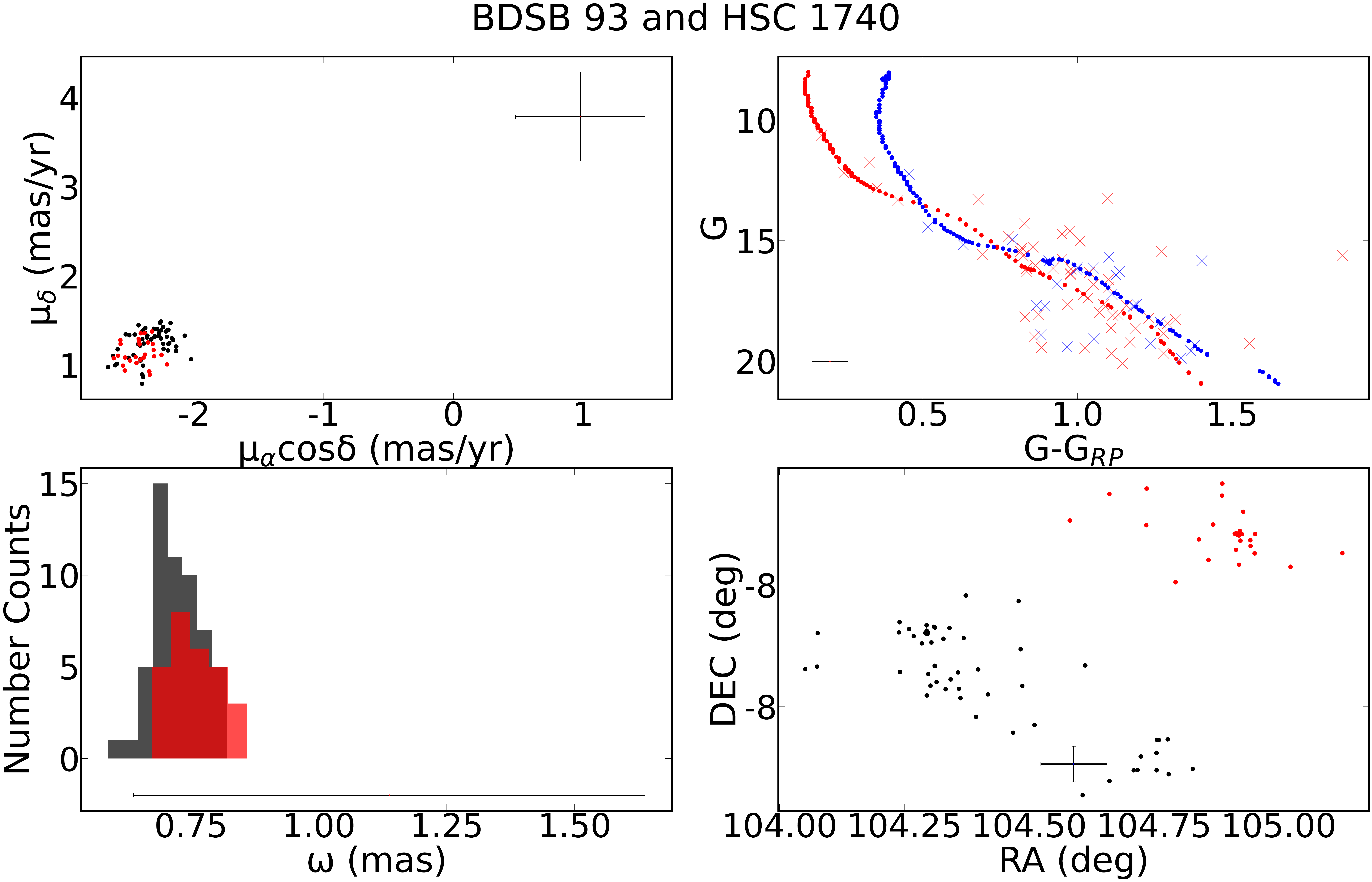}
    \includegraphics[width=0.49\hsize]{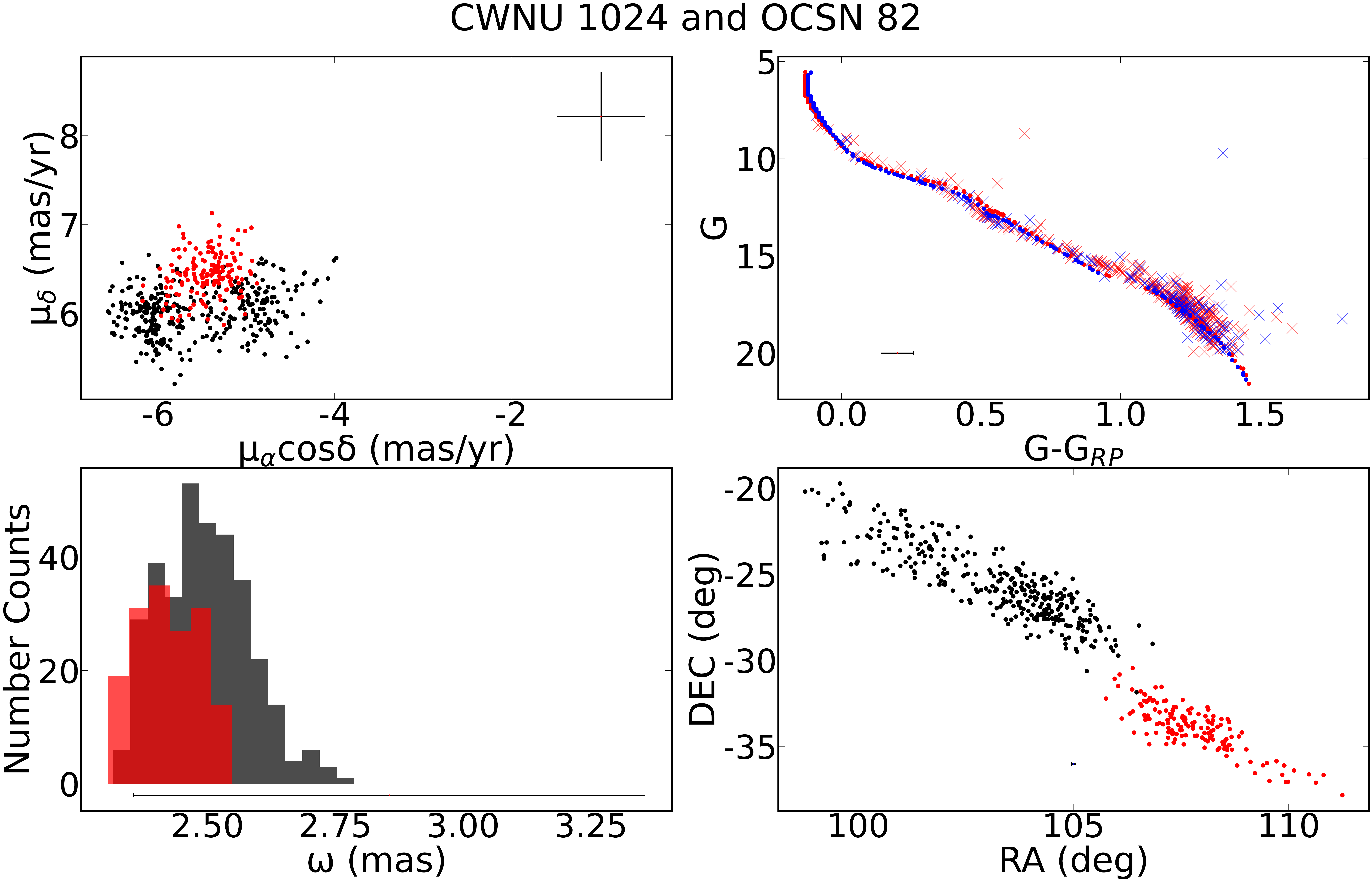}
    \includegraphics[width=0.49\hsize]{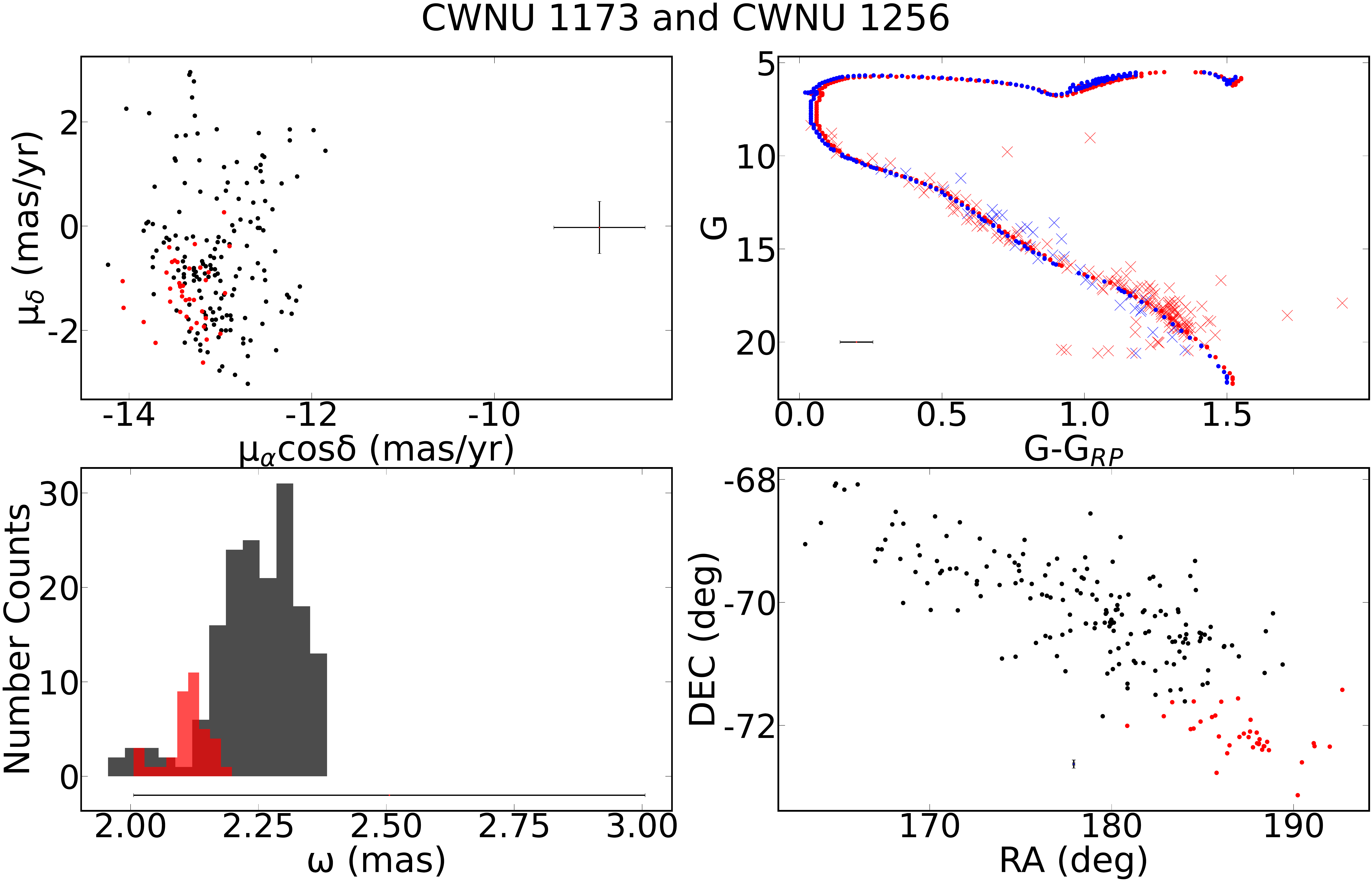}
    \includegraphics[width=0.49\hsize]{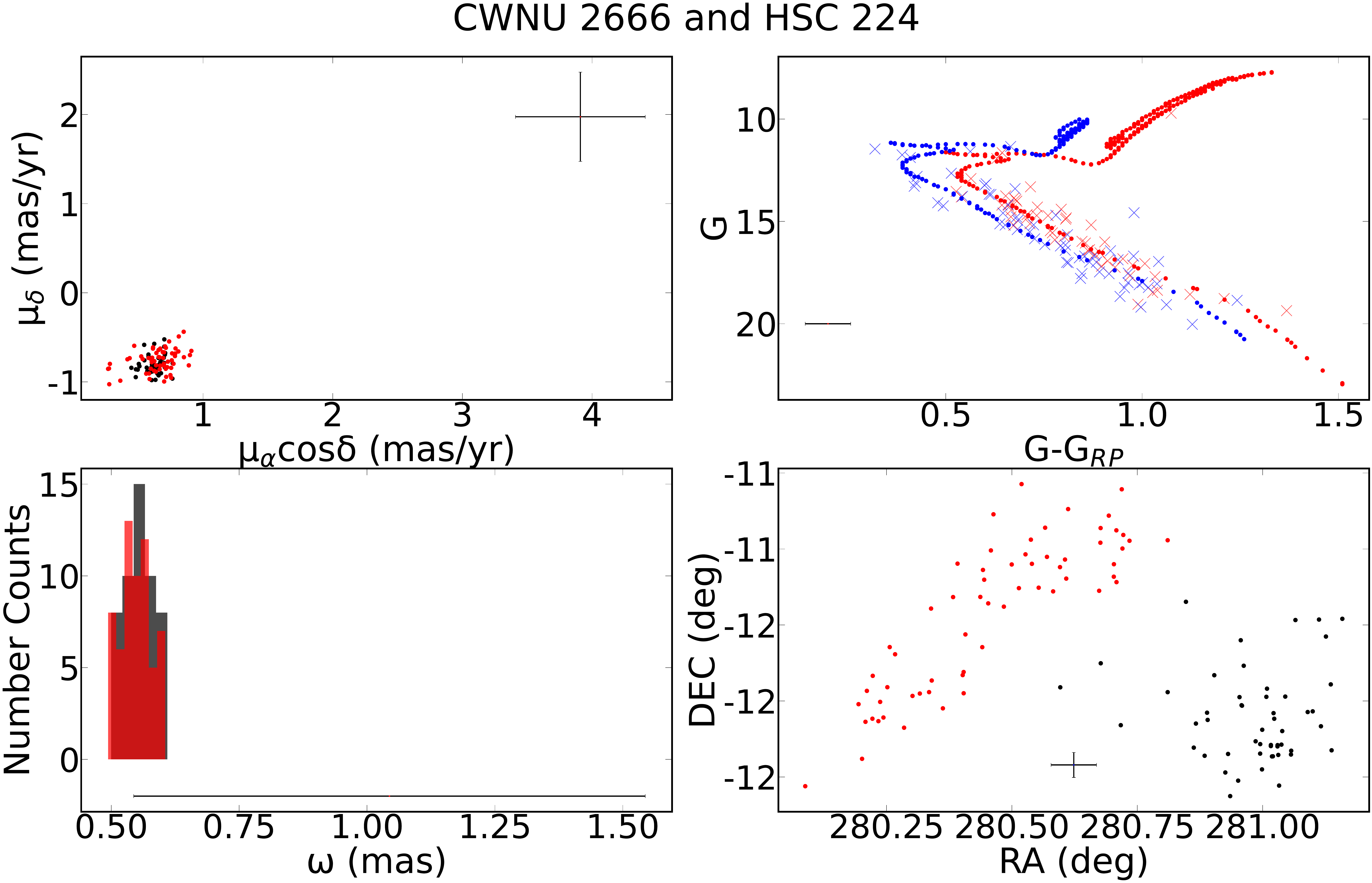}
    \includegraphics[width=0.49\hsize]{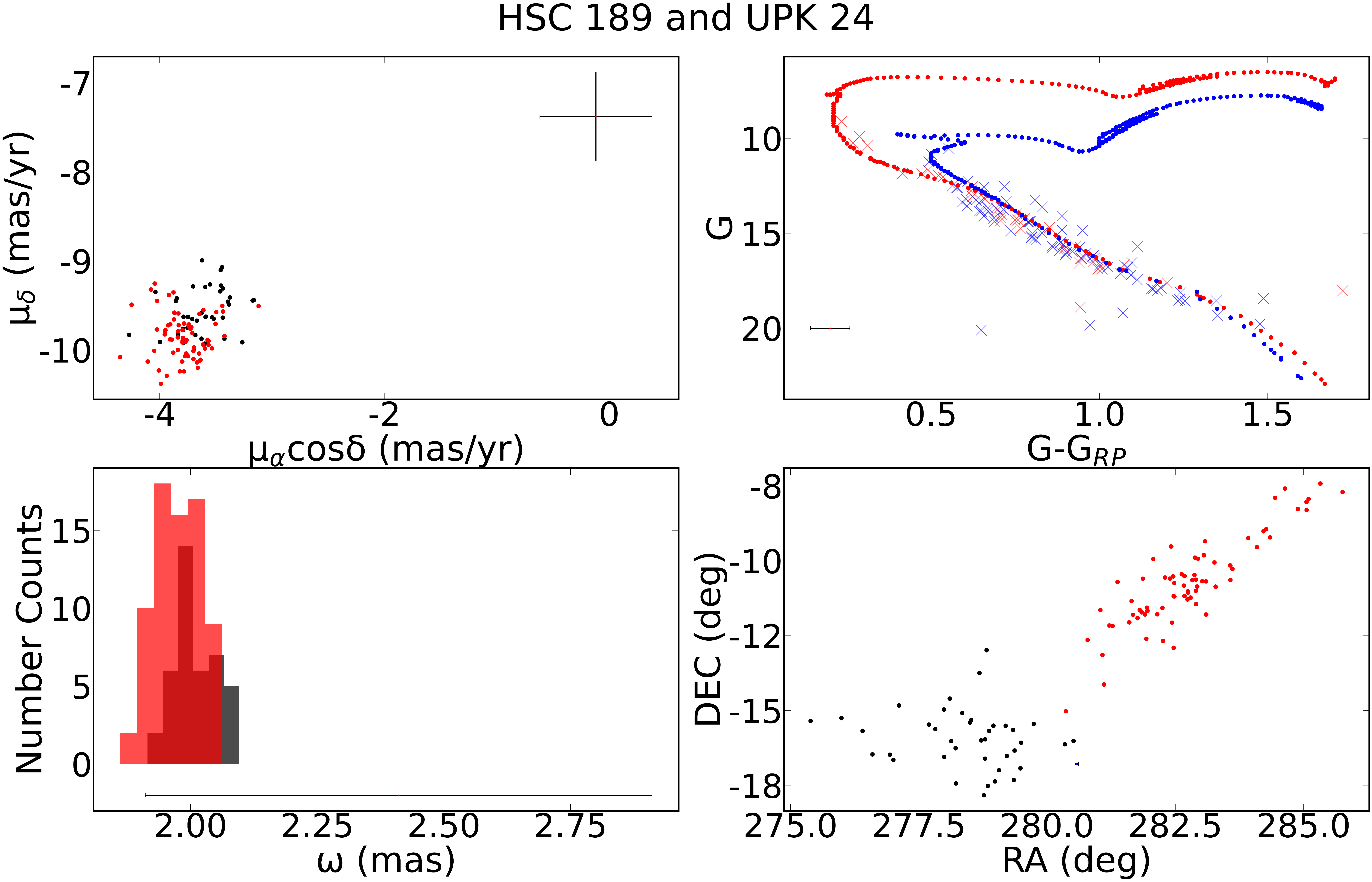}
    \includegraphics[width=0.49\hsize]{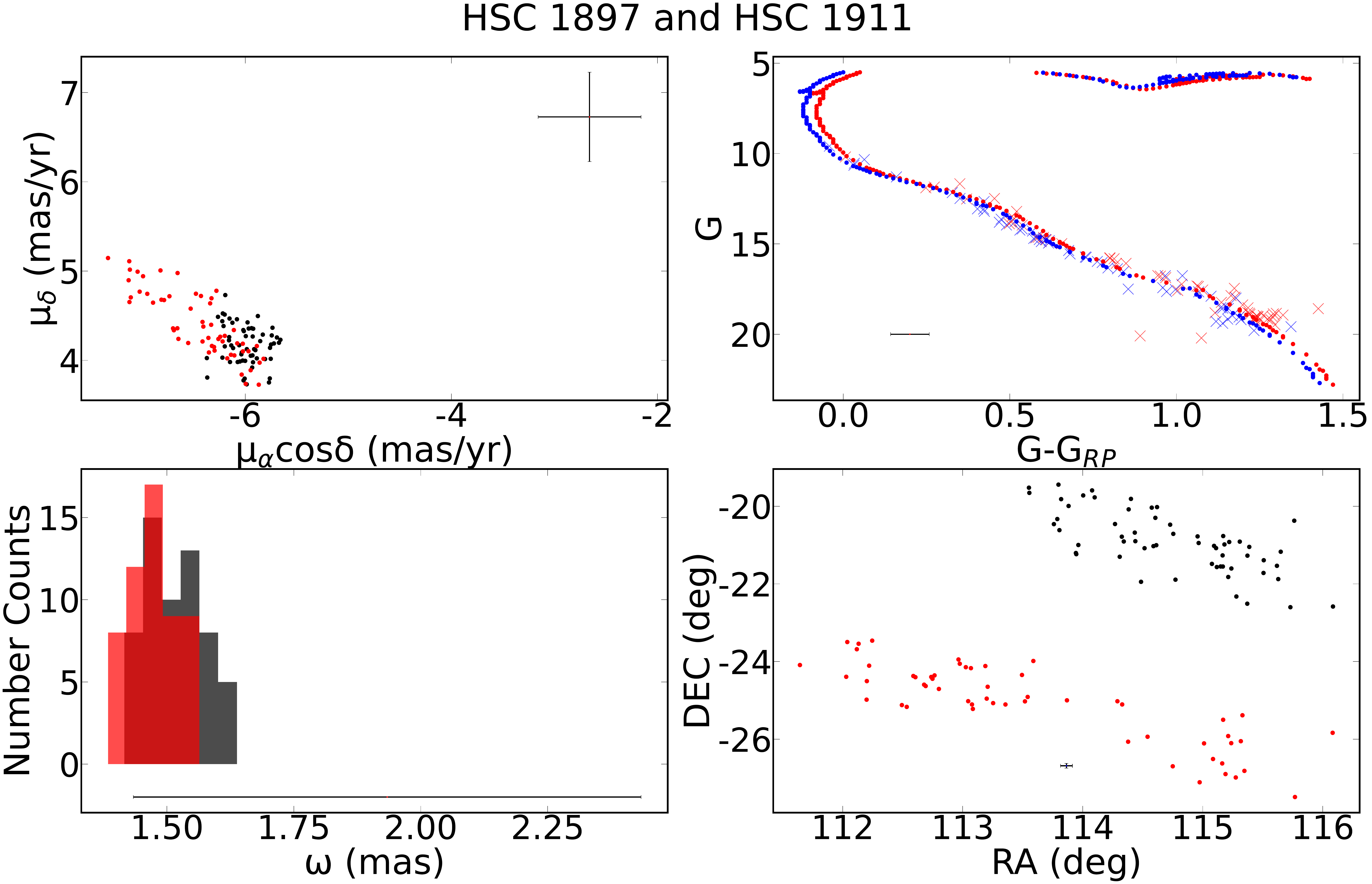}
    \includegraphics[width=0.49\hsize]{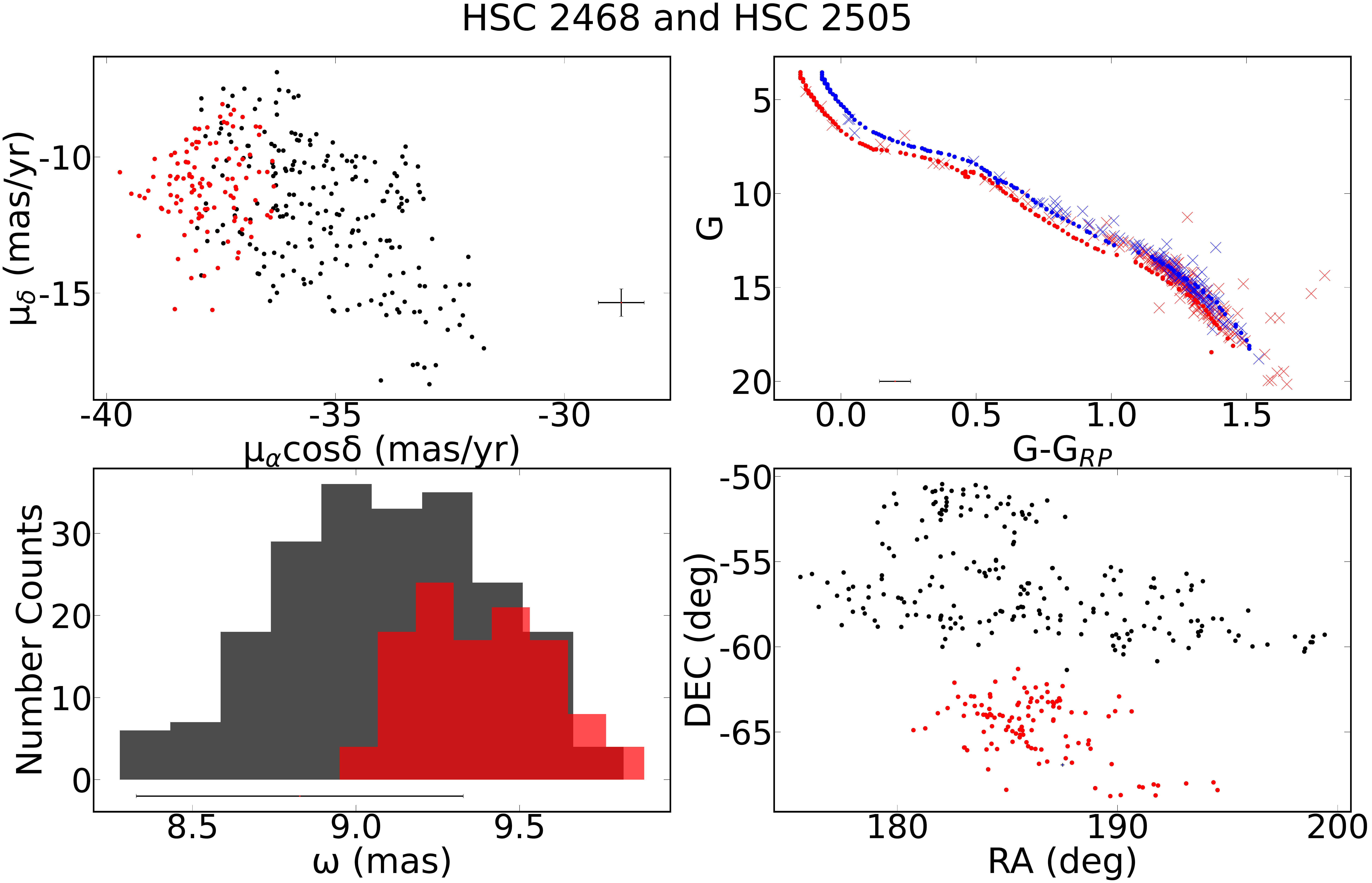}
    \caption{Distributions of PMs, CMDs, parallaxes, and celestial position of member stars of nine newly found candidate CBOCs. In the panels expect the CMD, black and red colours are for two sub-clusters respectively. For the CMD panel, blue and red colours are used, in which crosses and points are for observed stars and best-fit isochrones. Error bars show the median uncertainties (from the Gaia report) in proper motion, magnitudes and parallax at 20\,mag, and in the coordinates (from all member stars of a cluster).}
    \label{sec:fig.4}
\end{figure*}

\begin{figure}[h]
    \centering
    \ContinuedFloat
    \includegraphics[width=\hsize]{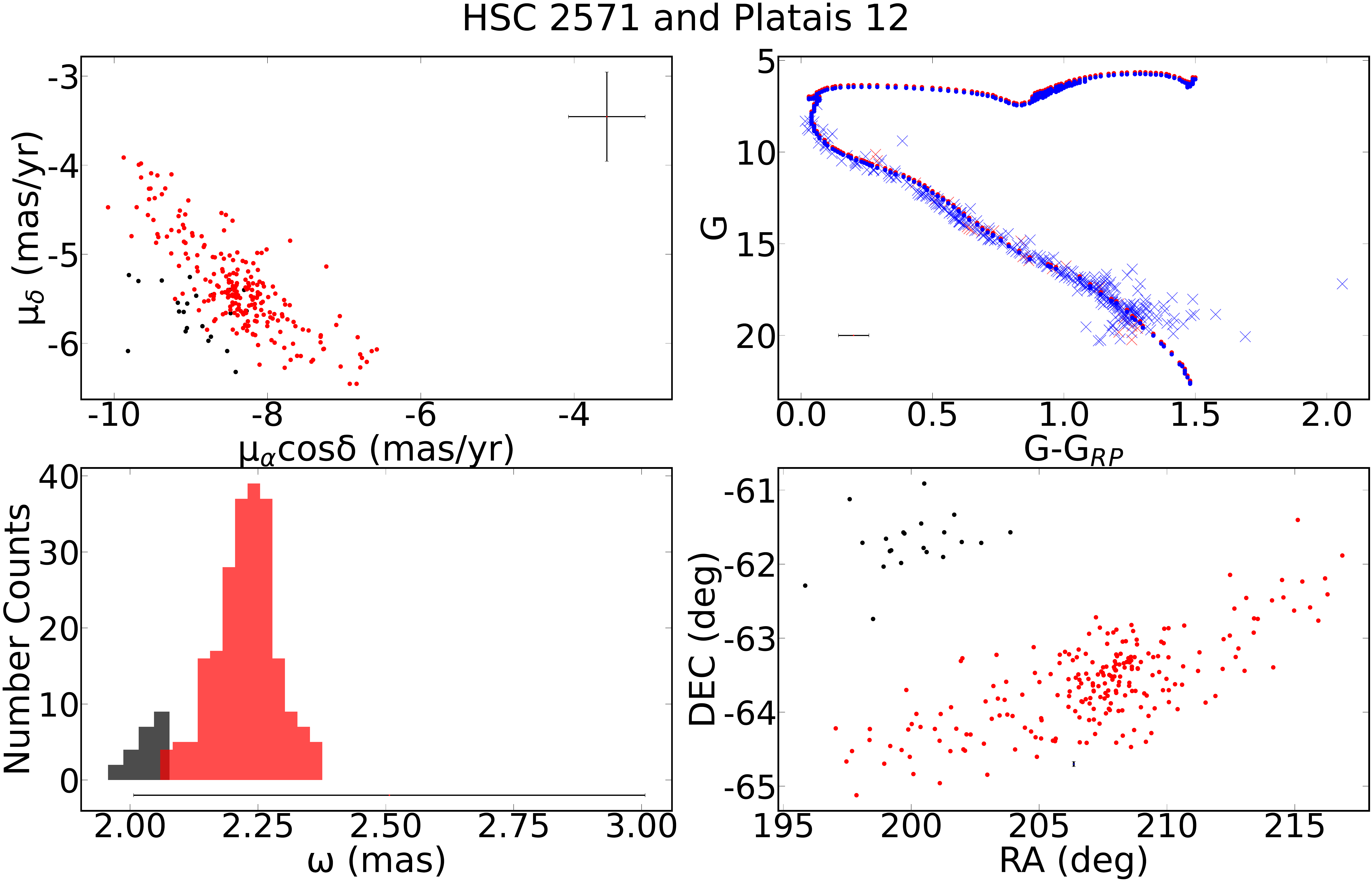}
    \caption{continued.}
\end{figure}

\section{Results} \label{sec:result}
\subsection{Newly found candidate CBOCs and their properties}

Via studying the similarity of spatial position and proper motions of the members of cluster pairs in HR23, dozens of rough candidate CBOCs are found.
Fig.~\ref{sec:fig.3} shows the classification of all cluster pairs. In the figure, blue points represent cluster pairs with separation less than the sum of tidal radii of member clusters but unsatisfying the proper motion criteria. Orange dots indicate cluster pairs that the two member clusters may belong to the same cluster as they are too close. Green points denote cluster aggregates. Red triangles indicate the discovered CBOC candidates by this work.
However, some of the candidate CBOCs have been found by previous works. The results are then cross-matched to the known list of binary clusters or binary cluster candidates \citep{1995AA...302...86S,1997A&AT...14..181L,2009AA...500L..13D,2017AA...600A.106C,2021AA...649A..54P}. If a member of a rough candidate CBOC has been listed as a sub-cluster of known binary cluster or candidate, the candidate CBOC is considered known. Many known binary clusters, e.g. Alessi 43 and Collinder 197, are removed from the list of newly discovered candidate CBOCs.
As a result, we identified 16 candidates for CBOCs. Three other candidates (HSC 633 and HSC 673, HSC 865 and HSC 958, HSC 1743 and Theia 267) are removed from the candidate CBOC list, because their member clusters HSC 633, HSC 958 and HSC 1743 are classified as moving groups rather than open clusters by HR23. We finally obtain 13 CBOC candidates. Table~\ref{sec:table.1} gives the spatial positions of them. The fundamental parameters of these CBOCs including colour excess, age, metallicity, and distance modulus are derived from fitting the CMDs to the PARSEC 1.2s isochrones \citep{2012MNRAS.427..127B} via a GPU version of Powerful CMD code \citep{2017RAA....17...71L}.
The code divides the CMD of star clusters into 1500 cells, including 50 colour bins and 30 magnitude bins, and uses a weight average difference (WAD) to assess the goodness of fit \citep{2017RAA....17...71L,2024RAA....24f5004D}. WAD is calculated using the formula:
\begin{equation}
   {\rm WAD} = \frac{\sum \omega_{i} |f_{\rm ob} - f_{\rm th}|}{\sum \omega_{i}},
\end{equation}
where $\omega_{i}$ is the weight of the $i$th cell, and $f_{\rm ob}$ and $f_{\rm th}$ are star fractions of observed and theoretical CMDs in the same cell respectively. $\omega_{i}$ relates to the stellar completeness of each cell, and it is set to 1 in this work. The parameters of stellar population models with the smallest WAD value are taken as the best-fit results.
In order to mitigate the effects of photometric uncertainty, we generate 20 CMDs randomly by taking the typical uncertainties in observational magnitudes into account and fit them, one by one. The result uncertainties are then calculated from the 20 series of best-fit parameters. This will be introduced in more detail later.
One can see these fundamental parameters in Table~\ref{sec:table.2}.

Fig.~\ref{sec:fig.4} compares the best fitting isochrones of two sub-clusters of 13 candidate CBOCs. The observed CMDs, proper motion distributions, parallax distributions, and two dimensional coordinates of these candidates are also shown. Note that the distance modulus, colour excess, age, and metallicity values of clusters are reported by Powerful CMD, and the parallax, proper motions, radial velocity (RV) and number of members (prob $>$ 0.7) are taken from HR23. In the panels except the CMDs, black and red colours are for two sub-clusters, respectively. For the CMDs, blue and red colours are used instead, in which crosses and points denote the observed stars and best-fit isochrones. We observe clearly that the results are consistent with CBOCs, which suggests that the method used by this work is reliable.

If the ages and metallicities of two sub-clusters are similar (age difference $<$ 50\,Myr and [Fe/H] difference $<$ 0.05\,dex), they may originated from the same molecular cloud. Such CBOC candidates can be assigned to PBOCs, and the candidates with large difference in the ages and metallicities should be excluded from PBOCs. When we check the age and metallicity differences of the 13 candidate CBOCs, nine of them are assigned to PBOC candidates, as we can see from Table~\ref{sec:table.2}. Note that the critical age and metallicity differences are set to slightly larger values compared to some previous works, because the uncertainties seem somewhat large in the CMD fitting.

\subsection{Deletion of candidates including moving group}

Although we eliminated sources classified as moving groups by HR23 at the beginning, there are still many objects that are more compatible with unbound moving groups in the OC sample \citep{2023AA...673A.114H}. A necessary procedure, therefore, is to demonstrate that two objects in a binary cluster candidate are gravitationally bound. We applied to the method of \cite{2019ApJ...870...32K} to the members of 13 candidate CBOCs.

This section calculates the one-dimensional velocity dispersions, $\sigma_{\rm 1D}$, and the velocity dispersions that are needed for virial equilibrium, $\sigma_{\rm vir}$, following the work of \cite{2019ApJ...870...32K}. The one-dimensional velocity dispersion is calculated by taking the mean variance of multidimensional velocity dispersions
\begin{equation}
   \sigma^2_{\rm 1D} = \frac{\sigma^2_{\rm pc1}+\sigma^2_{\rm pc2}}{2},
\end{equation}
where $\sigma_{\rm pc1}$ and $\sigma_{\rm pc2}$ are semi-major and semi-minor axes of the ellipse of proper motions. The 68.3 percent of member stars of a star cluster distribute in this ellipse, i.e. including stars that are within 1$\sigma$ error on proper motion. The directions of two components of velocity dispersion are calculated by principal component analysis (PCA). ``pc1'' and ``pc2'' denote the directions of the first and second components of PCA.

The velocity dispersion that is needed for virial equilibrium is calculated by the equation
\begin{equation}
    \sigma_{\rm vir} = \sqrt{\frac{G~Mass}{\eta~r_{\rm hm}}},
\end{equation}
where $G$ is the gravitational constant and $\eta$ is the mass profile parameter. A plummer model yields $\eta \approx$ 10 \citep{2010ARA&A..48..431P}, but many young clusters have $\eta < $10. In this case, we take two values (10 and 5) for $\eta$ in this work. $Mass$ and $r_{\rm hm}$ are the mass and half-mass radius of system respectively. We fit the observed CMD of a cluster to those of stellar populations with different IMFs to estimate stellar mass. The half-light radii ($r_{\rm hm}$) are calculated using the masses of HR23, because both the masses and positions of stars have been given. If a system has $\sigma_{\rm 1D} > \sqrt{2}\sigma_{\rm vir}$, its total energy would be positive and the system would be unbound. Thus we give the values for $\sqrt{2}\sigma_{\rm vir}$ in this work.

The results are shown in Table~\ref{sec:table.3}, in which the gravitational status of cluster candidates are also given. Objects with $\sigma_{\rm 1D} < \sqrt{2}\sigma_{\rm vir}$ are assigned as bound ones (``B''), while those with $\sigma_{\rm 1D} > \sqrt{2}\sigma_{\rm vir}$ are assigned as unbound ones (``U''), when taking both $\eta$ values. If a system has $\sqrt{2}\sigma_{\rm vir10}< \sigma_{\rm 1D} < \sqrt{2}\sigma_{\rm vir5}$, the object is thought as an uncertain one (``O'') (see e.g. ADS 16795, similar to the result of \citealt{2018MNRAS.481.3953A}). Four candidate CBOCs, i.e. (ADS 16795 and HSC 976), (HSC 477 and HSC 759), (HSC 1630 and HSC 1644), and (HSC 2204 and OC 0508), are excluded from the candidate CBOC list, because at least one member of them is possibly unbound according to the comparison of $\sigma_{\rm 1D}$ and $\sqrt{2}\sigma_{\rm vir}$.

\subsection{Identification of bound pairs}
In previous works \citep{2009AA...500L..13D,2022AA...666A..75S}, the authors determined whether cluster pairs constituted physically binary clusters based on parameters such as spatial proximity, and similarity in age, metallicity, proper motion, and parallax. However, such information is not enough for making a definitive confirmation, as we do not know which pairs are gravitationally bound. We therefore perform a deeper confirmation via a reliable method of \cite{2004ApJ...612..215M}. Some parameters, i.e. cluster mass, Roche radius, and orbital velocity are calculated for the confirmation.

\subsubsection{Cluster mass}
Gaia DR3 supplies important data, including proper motions, radial velocities of stars, and estimates the corresponding parameters of star clusters. However, there are still some challenges in estimating the masses of clusters, which are necessary for calculating the Roche radius and orbital velocity of star clusters. The two parameters are important to judge gravitationally bound and unbound cluster pairs. Many different methods for determining cluster mass are employed in researches. The main methods include the King profile method \citep{1962AJ.....67..471K}, the virial theorem \citep{1983A&A...118..361M}, and the integrated stellar luminosity function \citep{2008A&A...487..557P}.

In this work, we use the stellar population models with the Kroupa initial mass function (IMF) \citep{2001MNRAS.322..231K} to estimate the cluster masses and their errors. The stellar isochrones of PARSEC 1.2s \citep{2012MNRAS.427..127B}, which include 15 metallicities ($Z$ = 0.0001, 0.0002, 0.0005, 0.001, 0.002, 0.004, 0.006, 0.008, 0.01, 0.014, 0.017, 0.02, 0.03, 0.04, 0.06) and 1200 ages from 1 and 5996\,Myrs with a step of 5\,Myr, are used to build stellar populations. In order to estimate the masses of star clusters more reliably, we constructed a series of stellar population models with IMF slopes ($\Gamma$) covering the range from 1.6 to 2.4 with a step of 0.1. The masses of star clusters are derived by fitting the theoretically constructed CMDs of stellar population models to observational data. We obtain nine mass values for each cluster, corresponding to the nine IMF slopes. The mass includes that of all artificial stars with initial masses between 0.08 and 120\,$M_\odot$. Because this mass is the initial mass, the current mass is somewhat lower. Then the 1$\sigma$ error of mass of a cluster is calculated from the nine fitted mass values. Based on the stellar population models with the best-fit IMF slope, we fit the CMD of each cluster again. We generate 20 CMDs randomly by taking the typical uncertainties in observational magnitudes (0.006, 0.108 and 0.052\,mag for $G$, $G_{BP}$ and $G_{RP}$ respectively) into account and fit them using the theoretical stellar populations. We will get 20 series of best-fit parameters (including age) and the standard deviation (i.e. $\sigma$ value) for each parameter. These $\sigma$ values are then taken as parameter uncertainties. One can refer to \cite{2015ApJ...806..198W} for a similar application of various IMF slopes in the determination of cluster parameters. The masses of clusters are listed in Table~\ref{sec:table.4}.
A comparison of the ages and masses of clusters derived by this work with some recent literatures can be found in Appendix A.

\subsubsection{Roche radius and orbital velocity}
This section calculates the Roche radius ($R_{\rm roche}$) and orbital velocity ($V_{orb}$) of each cluster pair. We calculated $R_{\rm roche}$ and $V_{orb}$ using the same equations as in \cite{2004ApJ...612..215M}. For a cluster pair, the relevant size Roche radius can be calculated as:
\begin{equation}
    R_{\rm roche} = a[0.38+0.2log(m_{1}/m_{2})] ^{1/2},
\end{equation}
where $a$ is the semi-major orbital axis, while $m_1$ and $m_2$ are the masses of two sub-clusters. $a$ is roughly taken as the distance between the centres of two sub-clusters. The masses of clusters are taken from Table~\ref{sec:table.4}. We estimated the errors in the distances between cluster pairs and in the tidal radii based on the errors in RA, DEC, and $\omega$ provided by HR23. Then the errors in other parameters, e.g. that in $R_{\rm roche}$, are calculated from the error in distance, according to the error propagation formula. Note that the systematic errors are not taken into account here.

The Roche radii $R_{\rm roche}$ can be compared to the tidal radii to determine whether two sub-clusters are gravitationally bound. For a bound pair, the $R_{\rm roche}$ should be smaller than the tidal radius, for each sub-cluster.

If a cluster pair is bound, the expected orbital period in years ($P_{orb}$) is:
\begin{equation}
    P_{orb}=9.3\times10^7a^{3/2}(m_{1}+m_{2})^{-1/2},
\end{equation}
where $a$ is the semi-major axis in pc, and $m_1$ and $m_2$ are the masses of the two sub-clusters in $M_{\rm \odot}$.

The orbital velocity can be computed by:
\begin{equation}
    V_{orb}=2\pi a/P_{orb},
\end{equation}
where $a$ is in km and $P_{\rm orb}$ in s. This leads to $V_{\rm orb}$ values in \,km~s$^-$$^1$. Considering $\sin$$i$ and the orbital phase factor, $V_{\rm orb}$ should be the maximum velocity difference between two sub-clusters. Because HR23 reports the RVs of clusters, we compute the radial velocity difference of the sub-clusters in each pair, $\Delta$\,RV, and compare it with $V_{\rm orb}$.

\subsection{Analysis of nine candidate CBOCs}

In this section, we analyse the nine candidate CBOCs to seek for gravitationally bound cluster pairs, i.e. binary clusters.
The comparison of orbital velocity and radial velocity differences indicates that eight pairs except (CWNU 2666 and HSC 224) are gravitationally bound at a 1$\sigma$ level (see Table~\ref{sec:table.4}), because $\Delta$RV is less than $V_{\rm orb}$. Note that the 1$\sigma$ errors in the cluster parameters are taken into account in the comparisons.
However, the results need to be rechecked by comparing the Roche radius and tidal radius of each sub-cluster because of the large uncertainties in radial velocities. Moreover, we also analyse the metallicities and ages of sub-clusters of the nine candidate CBOCs. This is useful for the future studies on the formation of these pairs in deeper.

\subsubsection{ASCC 100 and HSC 534}
For the pair ASCC 100 and HSC 534, the difference in RV, i.e. $\Delta$RV, is 10.54\,$\pm$\,14.56\,km~s$^-$$^1$. It can be less than the orbital velocity $V_{\rm orb}$ (0.58\,$\pm$\,0.0011\,km~s$^-$$^1$) when the uncertainties are taken into account. However, the $R_{\rm roche}$ of sub-cluster HSC 534 (12.96\,$\pm$\,0.35\,pc) is larger than its tidal radius (10.39\,$\pm$\,0.09\,pc). This suggests that this pair is not bound. The CMDs of clusters ASCC 100 and HSC 534 share almost the same isochrone. Our results show that they share similar ages and metallicities. Because the tidal radius results from different works can be different a lot, ASCC 100 and HSC 534 is potentially a PBOC.

\subsubsection{BDSB 93 and HSC 1740}
For the pair BDSB 93 and HSC 1740, the $R_{\rm roche}$ of sub-cluster HSC 1740 (18.71\,$\pm$\,0.30\,pc) is larger than the tidal radius (7.64\,$\pm$\,0.32\,pc). It means that this pair is not a bound one. The age of HSC 1740 is 8\,Myr when taking the HR23 results, in agreement with our result of 13\,Myr. BDSB 93 is also a young OC with an age of 37\,Myr. If this pair is binary cluster, it will be a PBOC.

\subsubsection{CWNU 1024 and OCSN 82}
HR23 reported the RVs of CWNU 1024 and OCSN 82 as 21.59$\pm$12.65\,km~s$^-$$^1$ and 15.47$\pm$10.47\,km~s$^-$$^1$, respectively. \cite{2023ApJS..265...12Q} similarly reported the RV for OCSN 82 as 18.22\,km~s$^-$$^1$, which, slightly different from HR23 result. The estimated Roche radii (43.27\,$\pm$\,1.14\,pc for CWNU 1024 and 31.51\,$\pm$\,1.56\,pc) is less than the tidal radius (61.10\,$\pm$\,2.71\,pc for CWNU 1024 and 29.60\,$\pm$\,0.68\,pc OCSN 82) for both two sub-clusters when uncertainties are taken into account. The pair of CWNU 1024 and OCSN 82 is therefore likely to be a gravitationally bound system. In addition, the CMDs of CWNU 1024 and OCSN 82 overlap and share almost the same isochrone, suggesting that they are similar in age, metallicity, and distance modulus. These characteristics suggest that CWNU 1024 and OCSN 82 may be a PBOC.

\subsubsection{CWNU1173 and CWNU 1256}

For the pair of CWNU1173 and CWNU 1256, the Roche radius of sub-cluster CWNU 1256 (24.33\,$\pm$\,2.41\,pc) is greater than the tidal radius (13.88\,$\pm$\,1.34\,pc). This pair should not be bound. \cite{2022ApJS..262....7H} report the ages of CWNU 1173 and CWNU 1256 as 22\,Myr and 35\,Myr, respectively. The ages derived from the isochrone fitting in this work seem somewhat older (98\,Myr), but the difference is within acceptable limits when the age uncertainties are taken into account. This can also be listed as a candidate of PBOC, because of the large uncertainties (can be as large as 20\,pc) in the tidal radii of clusters.

\subsubsection{CWNU 2666 and HSC 224}
For the pair CWNU 2666 and HSC 224, the Roche radius of sub-cluster CWNU 2666 (15.12\,$\pm$\,2.08\,pc) is larger than its tidal radius (6.99\,$\pm$\,0.34\,pc). This pair seems not a bound one. HR23 reported ages of 98\,Myr and 64\,Myr for CWNU 2666 and HSC 224. \cite{2024AJ....167...12C} similarly reported their ages of 54\,Myr and 50\,Myr. \cite{2023ApJS..264....8H} reported the age of CWNU 2666 as 50\,Myr. In previous studies, their ages do not exceed 100\,Myr. However, the ages obtained by this study are older than 1\,Gyr. We scrutinised the CMDs of two clusters and found that the presence of a few stars in the upper right of the main sequence leads to the older ages. The CMD used in this work is different from previous works. In order to check the difference, we compared our results with \cite{2024AJ....167...12C}. Note that the $A_{\rm v}$ values are estimated by fitting relations $A_{\rm v}$ = 3.1 $E$(B - V) and $E$(G - G$_{RP}$) = 0.705 $E$(B - V) \citep{2018MNRAS.479L.102C}. The extinction, metallicity, and distance modulus of CWNU 2666 are 1.45\,mag, 0.0085 and 11.36\,mag, while those of HSC 224 are 0.87\,mag, 0.0085 and 10.68\,mag, respectively. The values reported by \cite{2024AJ....167...12C} are 2.86\,mag, 0.0246 and 11.18\,mag for CWNU 2666, and 2.21\,mag, 0.0459 and 11.51\,mag for HSC 224. We see that the extinctions and metallicities obtained by different works differ clearly. This suggests that the identification of member stars affects the stellar population parameters significantly.

\subsubsection{HSC 189 and UPK 24}
In the pair of HSC 189 and UPK 24, the Roche radius of HSC 189 (30.00\,$\pm$\,1.22\,pc) is larger than its tidal radius (24.62\,$\pm$\,0.67\,pc), suggesting that this pair is likely not bound. HR23 reports an age of 125\,Myr for HSC 189. The age of UPK 24 varies across different catalogues, with ages reported as 1047\,Myr \citep{2021MNRAS.504..356D}, 269\,Myr \citep{2020AA...640A...1C}, and 282\,Myr \citep{2022A&A...659A..59T}, with the oldest result exceeding 1\,Gyr. Based on the data reported in the catalogues, the ages of these two OCs differ significantly. The results of this work confirms this (see Table~\ref{sec:table.2}). The pair of HSC 189 and UPK 24 is certainly not a PBOC.

\subsubsection{HSC 1897 and HSC 1911}
For the pair HSC 1897 and HSC 1911, the Roche radius of HSC 1897 (30.05\,$\pm$\,1.02\,pc) is larger than its tidal radius (24.50\,$\pm$\,1.16\,pc). The two sub-clusters have similar metallicities and ages (see Table~\ref{sec:table.2}). This suggests that this is not a bound pair but a primordial pair.

\subsubsection{HSC 2468 and HSC 2505}
The pair HSC 2468 and HSC 2505 seems unbound, because the Roche radius of HSC 2505 (14.69\,$\pm$\,0.66\,pc) is larger than its tidal radius (5.48\,$\pm$\,0.17\,pc). This is an unbound pair. The ages and metallicities of two sub-clusters are similar, as we see from Table~\ref{sec:table.2}.

\subsubsection{HSC 2571 and Platais 12}
For the pair HSC 2571 and Platais 12, \cite{2017AstBu..72..257L} provided an RV of -12.0\,km~s$^-$$^1$ for cluster Platais 12, which is essentially in agreement with the HR23 result of -12.25 $\pm$ 8.16\,km s$^-$$^1$. HSC 2571 has an RV of -8.93\,$\pm$\,7.52\,km s$^-$$^1$. The RV difference suggests that this pair is possibly bound. However, the Roche radius of HSC 2571 (24.70\,$\pm$\,1.42\,pc) is larger than its tidal radius (14.33\,$\pm$\,0.64\,pc), suggesting that this pair is not gravitationally bound. The age of Platais 12 ranges from 79\,Myr \citep{2022ApJS..262....7H} to 169\,Myr \citep{2017AstBu..72..257L}. We obtained an age of 162\,Myr by CMD fitting, consistent with the results of \cite{2017AstBu..72..257L}. Both HSC 2571 and Platais 12 are young OCs of the same age. It means that this is a candidate of primordial but not a bound pair.

\begin{figure}[h]
    \centering
    \includegraphics[width=\hsize]{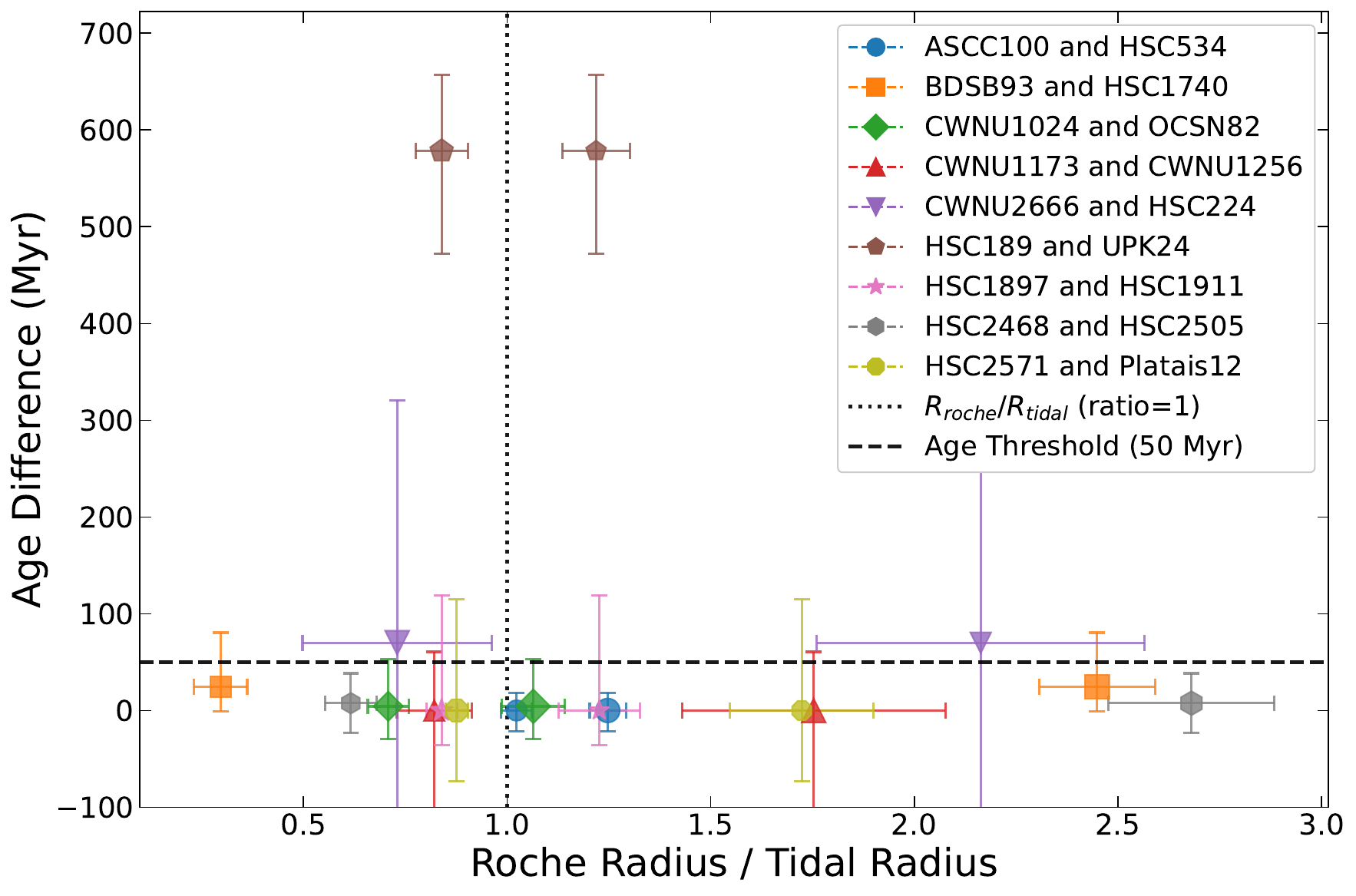}
    \caption{Distribution of cluster pairs in space of age difference versus ratio of $R_{\rm roche}$ to the tidal radius for each cluster.
    The sub-clusters of a pair are shown in the same colour and shapes. If two sub-clusters of a pair differ in age by less than 50\,Myr (black horizontal dotted line) and the ratio of $R_{\rm roche}$ to tidal radius of each sub-cluster is less than 1 (vertical black dotted line), it is assigned to a PBOC.}
    \label{sec:fig.5}
\end{figure}

\begin{figure}[h]
    \centering
    \includegraphics[width=\hsize]{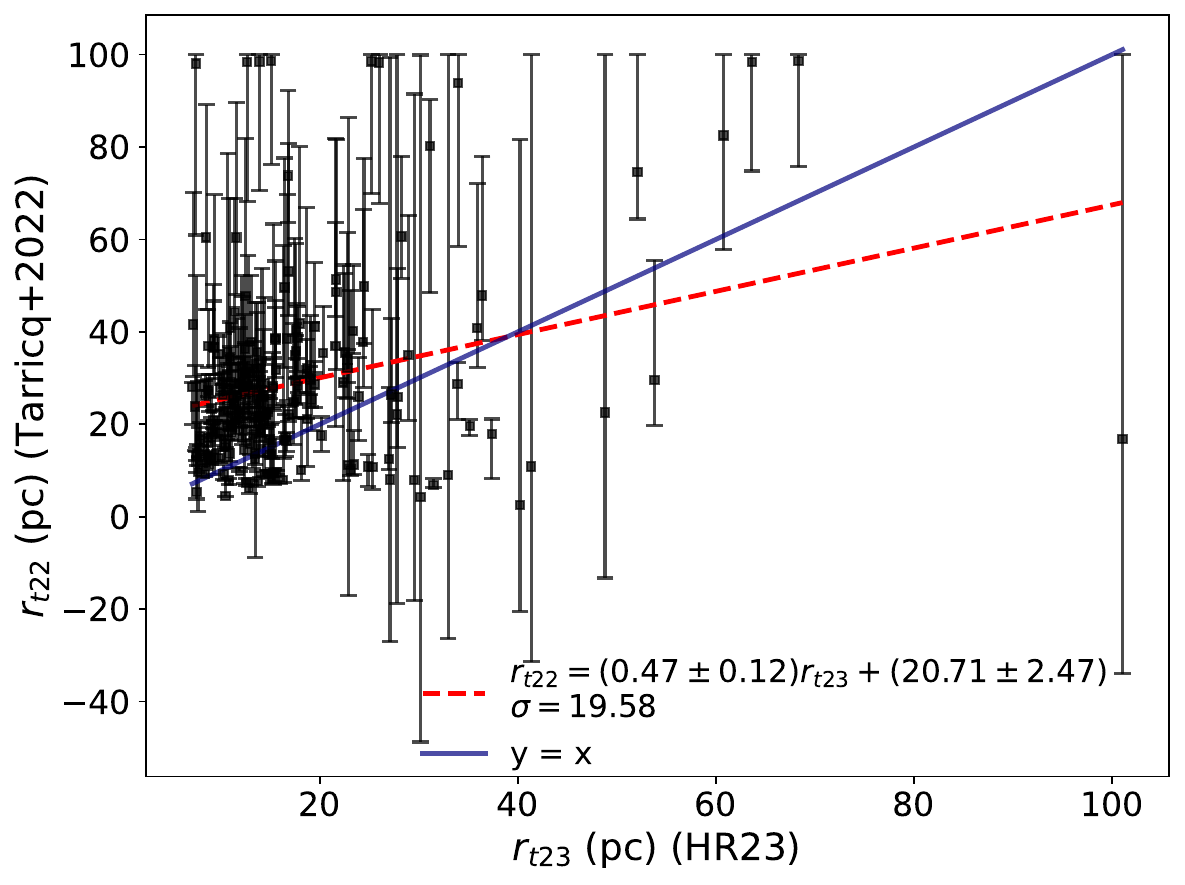}
    \caption{Comparison of tidal radii obtained by HR23 ($r_{\rm t23}$) and \cite{2022A&A...659A..59T} ($r_{\rm t22}$), for 202 clusters. The tidal radius errors are taken from \cite{2022A&A...659A..59T}. Blue line is the isoline, and red line is a linear fit to the correlation.}
    \label{sec:fig.6}
\end{figure}

In Fig.~\ref{sec:fig.5}, we show the distribution of cluster pairs in a space of age difference versus ratio of $R_{\rm roche}$ to the tidal radius for each cluster.
If a pair of clusters differ in age by less than 50 Myr (black horizontal dotted line) and the ratios of $R_{\rm roche}$ to the tidal radius are less than 1 (vertical black dotted line), the pair is sorted as PBOC candidate.
In summary, when the 1$\sigma$ uncertainties in mass, distance, and tidal radius are taken into account, a candidate cluster pair (CWNU 1024 and OCSN 82) is suggested to be bound binary cluster, at a level of 1$\sigma$. It remains unclear that the other eight pairs are gravitationally bound or not, because of the large uncertainties in tidal radii and radial velocities. In fact, the tidal radii of clusters in the catalogue HR23 are much less than other works, e.g. \cite{2022A&A...659A..59T}. \cite{2023AA...673A.114H} shows that the tidal radii of most clusters of a test sample of 202 clusters range from 20 to 100\,pc when taking the results of \cite{2022A&A...659A..59T}, but they range from 0 to 20\,pc when taking the results of HR23.
However, the catalogue of \cite{2022A&A...659A..59T} only includes one of the star clusters investigated in this study. In order to quantify the difference between the estimations of cluster tidal radii in two works, we compare the tidal radii of a sample of 202 clusters and fit their correlation using the least square method, as shown in Fig.~\ref{sec:fig.6}. If this loose correlation is used to correct the tidal radii of HR23, the values should increase by at least 17. It suggests that all nine cluster pairs in this work would be gravitationally bound.

\begin{table*}
\renewcommand{\arraystretch}{1.2}
    \centering
    \caption{Spatial positions of 13 newly discovered candidate CBOCs.}
    \begin{tabular}{ccccccccccccccccc}
    \hline
Pair	&	Name$_{\rm 1}$	&RA$_{\rm 1}$	&DEC$_{\rm 1}$ 	&	Parallax$_{\rm 1}$	&	$r_{\rm1}$	&	Name$_{\rm 2}$	&RA$_{\rm 2}$	&DEC$_{\rm 2}$ 	&	Parallax$_{\rm 2}$	&	$r_{\rm2}$	\\
        &                   &   [deg]           &   [deg]           &   [mas]               &   [pc]                &                   &    [deg]          &    [deg]          &
[mas]   &   [pc]   \\
    \hline
1	&	ADS16795	&	352.59	&	58.55	&	4.76$\pm$0.10	&	30.82$\pm$2.92	&	HSC976	&	10.00	&	79.32	&	6.04$\pm$0.46	&	63.79$\pm$4.81	\\
2	&	ASCC100	&	285.40	&	33.60	&	2.75$\pm$0.09	&	18.35$\pm$0.44	&	HSC534	&	290.15	&	34.15	&	2.67$\pm$0.03	&	10.39$\pm$0.09	\\
3	&	BDSB93	&	104.30	&	-8.20	&	0.72$\pm$0.05	&	17.24$\pm$0.14	&	HSC1740	&	104.92	&	-7.80	&	0.76$\pm$0.05	&	7.64$\pm$0.32	\\
4	&	CWNU1024	&	104.29	&	-27.39	&	2.50$\pm$0.09	&	61.10$\pm$2.71	&	OCSN82	&	108.53	&	-35.06	&	2.42$\pm$0.06	&	29.6$\pm$0.68	\\
5	&	CWNU1173	&	179.92	&	-70.29	&	2.24$\pm$0.08	&	34.80$\pm$1.66	&	CWNU1256	&	188.17	&	-72.30	&	2.12$\pm$0.04	&	13.88$\pm$1.34	\\
6	&	CWNU2666	&	281.02	&	-11.91	&	0.56$\pm$0.03	&	6.99$\pm$0.34	&	HSC224	&	280.23	&	-11.80	&	0.55$\pm$0.03	&	21.35$\pm$4.03	\\
7	&	HSC189	&	279.07	&	-17.00	&	2.01$\pm$0.04	&	24.62$\pm$0.67	&	UPK24	&	281.88	&	-11.70	&	1.98$\pm$0.04	&	40.23$\pm$1.79	\\
8	&	HSC477	&	234.70	&	36.62	&	6.36$\pm$0.17	&	22.91$\pm$0.19	&	HSC759	&	225.15	&	59.39	&	10.16$\pm$0.94	&	73.10$\pm$0.20	\\
9	&	HSC1630	&	89.35	&	0.10	&	6.88$\pm$1.26	&	71.92$\pm$0.91	&	HSC1644	&	79.40	&	-6.83	&	5.83$\pm$0.17	&	10.93$\pm$0.12	\\
10	&	HSC1897	&	115.15	&	-21.55	&	1.52$\pm$0.06	&	24.50$\pm$1.16	&	HSC1911	&	112.70	&	-24.40	&	1.47$\pm$0.05	&	36.40$\pm$0.44	\\
11	&	HSC2204	&	137.36	&	-50.39	&	0.41$\pm$0.03	&	11.30$\pm$5.05	&	OC0508	&	137.55	&	-49.79	&	0.39$\pm$0.04	&	13.49$\pm$0.96	\\
12	&	HSC2468	&	182.23	&	-51.54	&	9.08$\pm$0.32	&	23.88$\pm$1.38	&	HSC2505	&	185.02	&	-64.10	&	9.35$\pm$0.19	&	5.48$\pm$0.17	\\
13	&	HSC2571	&	199.17	&	-61.81	&	2.03$\pm$0.03	&	14.33$\pm$0.64	&	Platais12	&	207.92	&	-63.45	&	2.22$\pm$0.08	&	46.17$\pm$0.44	\\

    \hline
    \end{tabular}
    \label{sec:table.1}
    \tablefoot{The subscripts 1 and 2 indicate the first and second sub-clusters respectively. $r_{\rm1}$ and $r_{\rm2}$ mean tidal radii of two sub-clusters. The data are taken from HR23.}
\end{table*}

\begin{table*}
\renewcommand{\arraystretch}{1.2}
    \centering
    \caption{Fundamental parameters and astrometric information of 13 newly discovered candidate CBOCs. }
    \begin{tabular}{ccccccccccccccccc}
    \hline
Pair 	&	Cluster	&	m-M	&	E(G - G$_{\rm RP}$)	&	Age	&	[Fe/H]	&	$\mu${$_\alpha$}cos$\delta$	&	$\mu${$_\delta$}	&		N	&	Note	\\
	&		&	[mag]	&	[mag]	&	[Myr]	&		&	[mas/yr]	&	[mas/yr]	&		Prob$>$0.7	&		\\
    \hline
1	&	ADS16795	&	$7.33^{+0.42}_{-0.32}$	&	$0.27^{+0.02}_{-0.11}$	&	$98^{+14}_{-4}$	&	$0.0085^{+0.0185}_{-0.0037}$	&	18.66$\pm$1.30	&	2.85$\pm$1.26	&	41 	&		\\
	&	HSC976	&	$6.56^{+0.60}_{-0.54}$	&	$0.14^{+0.10}_{-0.11}$	&	$14^{+1}_{-1}$	&	$0.0152^{+0.0118}_{-0.0067}$	&	17.16$\pm$5.80	&	5.32$\pm$5.30	&	88 	&		\\
2	&	ASCC100	&	$7.71^{+0.41}_{-0.37}$	&	$0.12^{+0.09}_{-0.08}$	&	$31^{+18}_{-3}$	&	$0.0085^{+0.0067}_{-0.0037}$	&	1.73$\pm$0.46	&	-3.45$\pm$0.29	&	87 	&	PBOC	\\
	&	HSC534	&	$7.71^{+0.38}_{-0.33}$	&	$0.11^{+0.07}_{-0.06}$	&	$31^{+18}_{-11}$	&	$0.0085^{+0.0067}_{-0.0037}$	&	1.68$\pm$0.18	&	-3.25$\pm$0.32	&	18 	&	PBOC	\\
3	&	BDSB93	&	$9.68^{+0.38}_{-0.80}$	&	$0.34^{+0.15}_{-0.13}$	&	$37^{+56}_{-24}$	&	$0.0027^{+0.0021}_{-0.0012}$	&	-2.33$\pm$0.14	&	1.25$\pm$0.16	&	55 	&	PBOC	\\
	&	HSC1740	&	$11.69^{+0.30}_{-0.47}$	&	$0.61^{+0.13}_{-0.21}$	&	$13^{+7}_{-0}$	&	$0.0152^{+0.0118}_{-0.0067}$	&	-2.42$\pm$0.11	&	1.14$\pm$0.14	&	27 	&	PBOC	\\
4	&	CWNU1024	&	$8.71^{+0.34}_{-0.28}$	&	$0.08^{+0.09}_{-0.06}$	&	$21^{+32}_{-5}$	&	$0.0481^{+0}_{-0.0211}$	&	-5.55$\pm$0.62	&	6.02$\pm$0.26	&	336 	&	PBOC	\\
	&	OCSN82	&	$8.60^{+0.25}_{-0.36}$	&	$0.09^{+0.10}_{-0.06}$	&	$26^{+48}_{-10}$	&	$0.0270^{+0.0211}_{-0.0118}$	&	-5.44$\pm$0.27	&	6.46$\pm$0.24	&	157 	&	PBOC	\\
5	&	CWNU1173	&	$8.56^{+0.43}_{-0.44}$	&	$0.17^{+0.12}_{-0.11}$	&	$98^{+116}_{-61}$	&	$0.0481^{+0}_{-0.0211}$	&	-13.03$\pm$0.43	&	-0.62$\pm$1.19	&	165 	&	PBOC	\\
	&	CWNU1256	&	$8.54^{+0.34}_{-0.33}$	&	$0.15^{+0.09}_{-0.08}$	&	$98^{+4}_{-4}$	&	$0.0481^{+0}_{-0.0211}$	&	-13.38$\pm$0.26	&	-1.33$\pm$0.56	&	37 	&	PBOC	\\
6	&	CWNU2666	&	$9.91^{+0.47}_{-0.36}$	&	$0.33^{+0.13}_{-0.13}$	&	$1480^{+220}_{-200}$	&	$0.0085^{+0.0067}_{-0.0037}$	&	0.61$\pm$0.07	&	-0.80$\pm$0.10	&	51 	&		\\
	&	HSC224	&	$9.81^{+0.44}_{-0.54}$	&	$0.20^{+0.18}_{-0.10}$	&	$1410^{+70}_{-120}$	&	$0.0085^{+0.0067}_{-0.0037}$	&	0.66$\pm$0.15	&	-0.76$\pm$0.13	&	61 	&		\\
7	&	HSC189	&	$9.25^{+0.41}_{-0.24}$	&	$0.32^{+0.08}_{-0.10}$	&	$98^{+88}_{-44}$	&	$0.0481^{+0}_{-0.0211}$	&	-3.62$\pm$0.23	&	-9.60$\pm$0.25	&	40 	&		\\
	&	UPK24	&	$9.25^{+0.44}_{-0.24}$	&	$0.31^{+0.12}_{-0.11}$	&	$676^{+65}_{-59}$	&	$0.0481^{+0}_{-0.0211}$	&	-3.80$\pm$0.20	&	-9.85$\pm$0.25	&	72 	&		\\
8	&	HSC477	&	$7.66^{+0.39}_{-0.53}$	&	$0.13^{+0.13}_{-0.04}$	&	$14^{+17}_{-1}$	&	$0.0481^{+0}_{-0.0211}$	&	-15.40$\pm$1.86	&	-6.18$\pm$1.95	&	24 	&		PBOC\\
	&	HSC759	&	$5.69^{+0.69}_{-0.48}$	&	$0.11^{+0.14}_{-0.09}$	&	$102^{+93}_{-80}$	&	$0.0481^{+0}_{-0.0329}$	&	-16.19$\pm$2.78	&	-3.64$\pm$2.92	&	152 	&		PBOC\\
9	&	HSC1630	&	$6.00^{+0.37}_{-0.27}$	&	$0.06^{+0.07}_{-0.06}$	&	$62^{+195}_{-31}$	&	$0.0270^{+0.0211}_{-0.0185}$	&	-16.43$\pm$2.12	&	-11.61$\pm$4.72	&	94 	&	PBOC	\\
	&	HSC1644	&	$7.75^{+0.41}_{-0.29}$	&	$0.09^{+0.07}_{-0.06}$	&	$14^{+1}_{-1}$	&	$0.0481^{+0}_{-0.0211}$	&	-15.69$\pm$0.99	&	-8.22$\pm$0.76	&	14 	&	PBOC	\\
10	&	HSC1897	&	$9.53^{+0.38}_{-0.37}$	&	$0.09^{+0.07}_{-0.07}$	&	$45^{+26}_{-20}$	&	$0.0481^{+0}_{-0.0329}$	&	-5.99$\pm$0.18	&	4.18$\pm$0.22	&	59 	&	PBOC	\\
	&	HSC1911	&	$9.53^{+0.41}_{-0.42}$	&	$0.05^{+0.07}_{-0.04}$	&	$45^{+117}_{-24}$	&	$0.0481^{+0}_{-0.0211}$	&	-6.39$\pm$0.35	&	4.35$\pm$0.32	&	55 	&	PBOC	\\
11	&	HSC2204	&	$11.71^{+0.78}_{-0.42}$	&	$0.86^{+0.19}_{-0.22}$	&	$741^{+279}_{-65}$	&	$0.0027^{+0.0021}_{-0.0012}$	&	-5.54$\pm$0.11	&	5.66$\pm$0.14	&	24 	&		\\
	&	OC0508	&	$11.81^{+1.18}_{-1.02}$	&	$0.87^{+0.20}_{-0.16}$	&	$813^{+78}_{-72}$	&	$0.0027^{+0.0058}_{-0.0018}$	&	-5.57$\pm$0.14	&	5.58$\pm$0.12	&	81 	&		\\
12	&	HSC2468	&	$6.00^{+0.37}_{-0.54}$	&	$0.08^{+0.13}_{-0.05}$	&	$13^{+30}_{-0}$	&	$0.0481^{+0}_{-0.0329}$	&	-35.27$\pm$1.51	&	-11.96$\pm$2.35	&	210 	&	PBOC	\\
	&	HSC2505	&	$5.31^{+0.41}_{-0.38}$	&	$0.13^{+0.09}_{-0.09}$	&	$20^{+31}_{-8}$	&	$0.0481^{+0}_{-0.0211}$	&	-37.79$\pm$0.81	&	-10.81$\pm$1.26	&	113 	&	PBOC	\\
13	&	HSC2571	&	$8.48^{+0.46}_{-0.36}$	&	$0.12^{+0.07}_{-0.08}$	&	$162^{+42}_{-64}$	&	$0.0481^{+0}_{-0.0329}$	&	-8.97$\pm$0.45	&	-5.62$\pm$0.27	&	22 	&	PBOC	\\
	&	Platais12	&	$8.58^{+0.50}_{-0.51}$	&	$0.12^{+0.09}_{-0.10}$	&	$162^{+95}_{-60}$	&	$0.0481^{+0}_{-0.0329}$	&	-8.31$\pm$0.77	&	-5.41$\pm$0.57	&	225 	&	PBOC	\\

    \hline
    \end{tabular}
    \label{sec:table.2}
    \tablefoot{PBOC means candidate of primordial binary open cluster. The $\mu${$_\alpha$}cos$\delta$ and $\mu${$_\delta$} are taken from HR23, while m-M, E(G -G$_{\rm RP}$), Age and [Fe/H] are determined by this work.}
\end{table*}

\begin{sidewaystable}
\renewcommand{\arraystretch}{1.2}
    \centering
    \caption{One-dimensional velocity dispersions ($\sigma_{1D}$) and velocity dispersions needed for virial equilibrium ($\sigma_{\rm vir10}$ and $\sigma_{\rm vir5}$) of 26 cluster candidates. }
    \begin{tabular}{cccccccccccccccccccccc}
    \hline
Cluster	    &   $\sigma_{1D}$	&   $E\sigma_{1D}$    &  $\sqrt{2}\sigma_{\rm vir10}$	&  $\sqrt{2}\sigma_{\rm vir5}$  & Status &&&	Cluster	    &   $\sigma_{1D}$	&   $E\sigma_{1D}$ & $\sqrt{2}\sigma_{\rm vir10}$	&    $\sqrt{2}\sigma_{\rm vir5}$  & Status\\
	        &	[km/s]	        &	[km/s]	&	[km/s]     &	[km/s]		                &&&&	            &    [km/s]	        &	[km/s]	&	[km/s]	   &	[km/s]	   &          \\
    \hline
ADS16795	&	1.35 	&	0.12 	&	1.11 	&	1.57 	&	O  &&& HSC976	&	4.19 	&	0.07 	&	1.94 	&	2.74 	&	U	\\
ASCC100	&	0.71 	&	0.13 	&	1.63 	&	2.31 	&	B &&& HSC534	&	0.47 	&	0.21 	&	1.02 	&	1.44 	&	B \\
BDSB93	&	1.35 	&	0.84 	&	0.85 	&	1.20 	&	B &&& HSC1740	&	1.08 	&	0.94 	&	0.60 	&	0.84 	&	B \\
CWNU1024	&	0.89 	&	0.14 	&	2.62 	&	3.70 	&	B &&&  OCSN82	&	0.55 	&	0.19 	&	2.30 	&	3.25 	&	B	\\
CWNU1173	&	2.07 	&	0.30 	&	2.00 	&	2.82 	&	B	&&& CWNU1256	&	1.15 	&	0.24 	&	1.49 	&	2.11 	&	B \\
CWNU2666	&	0.88 	&	0.56 	&	0.77 	&	1.10 	&	B &&& HSC224	&	1.21 	&	0.69 	&	1.04 	&	1.48 	&	B \\
HSC189	&	0.50 	&	0.09 	&	1.49 	&	2.10 	&	B &&& UPK24	&	0.65 	&	0.20 	&	1.72 	&	2.44 	&	B	\\
HSC477	&	1.37 	&	0.04 	&	0.97 	&	1.37 	&	O	&&& HSC759	&	1.36 	&	0.04 	&	2.44 	&	3.45 	&	B	\\
HSC1630	&	2.55 	&	0.06 	&	0.90 	&	1.28 	&	U	&&& HSC1644	&	0.69 	&	0.07 	&	0.71 	&	1.01 	&	B \\
HSC1897	&	0.81 	&	0.37 	&	1.13 	&	1.60 	&	B &&& HSC1911	&	0.95 	&	0.18 	&	0.82 	&	1.16 	&	B	\\
HSC2204	&	4.89 	&	1.33 	&	0.50 	&	0.71 	&	U	&&& OC0508	&	7.98 	&	1.26 	&	0.83 	&	1.18 	&	U \\
HSC2468	&	1.00 	&	0.02 	&	2.96 	&	4.19 	&	B	&&& HSC2505	&	0.54 	&	0.02 	&	2.94 	&	4.16 	&	B \\
HSC2571	&	0.92 	&	0.23 	&	0.96 	&	1.36 	&	B &&& Platais12	&	1.55 	&	0.15 	&	2.45 	&	3.47 	&	B	\\
    \hline
    \end{tabular}
    \label{sec:table.3}
    \tablefoot{$\sigma_{\rm vir10}$ and $\sigma_{\rm vir5}$ denote the  velocity dispersions needed for virial equilibrium that are calculated by taking 10 and 5 for mass profile parameter $\eta$. $E\sigma_{1D}$ is the uncertainty in $\sigma_{1D}$. ``B'', ``U'' and ``O'' denote bound, unbound and uncertain systems, respectively. }
\end{sidewaystable}

\begin{sidewaystable}
\renewcommand{\arraystretch}{1.2}
    \centering
    \caption{Parameters for checking gravitationally bound and unbound cluster pairs.}
    \begin{tabular}{llllllllllllllllllllll}
    \hline
Name$_{\rm 1}$	&	Name$_{\rm 2}$	&	$Mass_{\rm 1}$	&	$Mass_{\rm 2}$ &	$r_{\rm 1}$	&	$R_{\rm roche1}$  &	$r_{\rm 2}$	&	$R_{\rm roche2}$		&	$\Delta$ $RV$	&	$V_{\rm orb}$	\\
	&		&	[M$_\odot$]	&	[M$_\odot$]	&	[pc]	& [pc]	& [pc]	& [pc]	&	[km~s$^-$$^1$]	&	[km~s$^-$$^1$]		\\
\hline
ASCC100	&HSC534	  &1664.59$\pm$118.79	&352.19$\pm$9.22	&18.35$\pm$0.44	  &18.78$\pm$0.24    &10.39$\pm$0.09	  &12.96$\pm$0.35	  &10.54$\pm$14.65         &0.58$\pm$0.0011	\\
BDSB93	&HSC1740  &416.29$\pm$10.45	&18027.73$\pm$2279.65	&17.24$\pm$0.14	  & 5.11$\pm$1.09    &7.64$\pm$0.32	  &18.71$\pm$0.30         &5.08$\pm$18.87	   &1.90$\pm$0.0059	\\
CWNU1024&OCSN82	  &7425.00$\pm$988.68	&1939.08$\pm$134.03	&61.10$\pm$2.71	  &43.27$\pm$1.14    &29.6$\pm$0.68	  &31.51$\pm$1.56	  &6.12$\pm$16.42	   &0.82$\pm$0.0012	\\
CWNU1173&CWNU1256 &1864.56$\pm$266.66	&928.03$\pm$268.96	&34.80$\pm$1.66	  &28.58$\pm$1.82    &13.88$\pm$1.34	  &24.33$\pm$2.14	  &1.49$\pm$13.05	   &0.53$\pm$0.0023	\\
CWNU2666&HSC224	  &713.96$\pm$104.22	&816.00$\pm$461.53	&6.99$\pm$0.34	  &15.12$\pm$2.08    &21.35$\pm$4.03	  &15.59$\pm$2.02	  &$---$	           &0.52$\pm$0.0041	\\
HSC189	&UPK24	  &621.52$\pm$50.97	&1040.68$\pm$138.86	&24.62$\pm$0.67	  &30.00$\pm$1.22    &40.23$\pm$1.79	  &33.77$\pm$1.08	  &7.58$\pm$20.17	   &0.37$\pm$0.0009	\\
HSC1897	&HSC1911  &1421.53$\pm$201.28	&1527.33$\pm$54.87	&24.50$\pm$1.16	  &30.05$\pm$1.02    &36.40$\pm$0.44	  &30.55$\pm$1.00	  &0.01$\pm$18.21	   &0.51$\pm$0.0007	\\
HSC2468	&HSC2505  &4984.30$\pm$861.13	&4930.71$\pm$464.42	&23.88$\pm$1.38	  &14.72$\pm$0.66    &5.48$\pm$0.17	  &14.69$\pm$0.66	  &1.12$\pm$12.06	   &1.35$\pm$0.0013	\\
HSC2571 &Platais12&550.45$\pm$73.75	&4059.79$\pm$116.79	&14.33$\pm$0.64	  &24.70$\pm$1.42    &46.17$\pm$0.44	  &40.45$\pm$0.87	  &3.32$\pm$11.10	   &0.61$\pm$0.0017	\\
    \hline
    \end{tabular}
    \label{sec:table.4}
    \tablefoot{$R_{\rm roche}$ is the Roche radius. $\Delta$$RV$ denotes the difference of radial velocities of two sub-clusters. $V_{\rm orb}$ denotes orbital velocity. Tidal radii $r_{\rm 1}$ and $r_{\rm 2}$ are taken from HR23, and the other parameters are calculated by this work. The boldface indicates gravitationally bound binary clusters.}
\end{sidewaystable}

\section{Conclusion and discussion}\label{sec:conclusion and discussion}

We utilized the three dimensional spatial data, two dimensional proper motion data, and colour-magnitude data of the OCs from a recently published cluster catalogue (HR23), to conduct a comprehensive search for close binary open clusters (CBOCs) in the Milky Way. The member stars of a close binary star cluster are thought to have similar proper motions and radial velocities. A genuine binary cluster must satisfy two conditions: (1) each sub-cluster must be gravitationally bound, and (2) the pair must be gravitationally bound. The methods of \cite{2004ApJ...612..215M} and \cite{2019ApJ...870...32K} were utilized to find out bound clusters and bound cluster pairs.
We ultimately identified nine candidate CBOCs from the HR23 catalogue, in which one pair is suggested to be a certain primordial binary cluster by the data.
Some of the other eight pairs are also potential bound pairs when considering alternative tidal radius determinations (e.g. \citealt{2022A&A...659A..59T}).
We also fit the CMDs of thirteen candidate CBOCs using the PARSEC 1.2s isochrones and Powerful CMD code, to determine the fundamental parameters such as distance modulus, metallicity, colour excess, and age. The masses of these clusters were also estimated by comparing the CMDs of star clusters and theoretical stellar populations with different IMF slopes.
The Roche radii are then calculated based on the distances between sub-clusters and their masses.

It is found that seven of candidate CBOCs in Table~\ref{sec:table.4} (the ones excluding CWNU2666 and HSC224, HSC189 and UPK224) formed simultaneously, i.e. PBOCs. This is because the two sub-clusters of such a binary cluster have similar stellar distribution, stellar proper motion distribution, stellar radial velocity distribution, metallicities, ages, and reddenings. For the other two pairs, i.e. (CWNU2666 and HSC224) and (HSC189 and UPK224), the sub-cluster ages are either too old or too different for a binary cluster system to remain stable over such a long period. This suggests that these two pairs possbily formed through tidal capture. The list of candidate binary clusters will be useful for numerous detailed studies, such as investigating the interactions between sub-clusters of binary star clusters or higher multiplicities.

In addition, via the method of \cite{2004ApJ...612..215M}, we calculated the orbital periods and velocities of cluster pairs, because the orbital velocities represent the maximum difference in radial velocities of cluster pairs. However, the large errors in radial velocities provided by Gaia make it difficult to determine whether a pair is a binary cluster based on orbital velocities alone. This is because no matter the value of the orbital velocities, $\Delta$RV can always be less than the orbital velocity due to the large error of the radial velocity. More precise radial velocity observations are therefore essential for confirming binary clusters.

In this work, we compare the kinetic and potential energies of OCs to distinguish between bound clusters and unbound moving groups. Four pairs of candidate binary clusters are finally excluded from bound clusters, as at least one object in such pair is an unbound system. However, the results actually depend on the method used for the study. \cite{2024A&A...686A..42H} calculated the size of the Roche surface of clusters (their Jacobi radius) to differentiate between bound and unbound clusters.
We checked our classification results with \cite{2024A&A...686A..42H} regarding these 13 pairs of clusters.
\cite{2024A&A...686A..42H} classified most of these clusters as unbound systems, retaining only three candidate CBOCs (BDSB 93 and HSC 1740, CWNU 2666 and HSC 224, HSC 2204 and OC 0508).
The classification of the cluster pair HSC 2204 and OC 0508 differs between this work and \cite{2024A&A...686A..42H}. Thus the results of this work require further investigations.

This paper studied only the candidate CBOCs in the Milky Way. However, there should be many more wide binary open clusters (WBOCs). We will study them in another work. This will be important for understanding the properties, formation and evolution of binary star clusters.

Since our research is based on the catalogue of \cite{2023AA...673A.114H}, the results actually depend on the accuracy of the data in the catalogue. There are usually some uncertainties in such catalogue. There is another catalogue that contains much more star clusters, i.e. UCC \citep{2023MNRAS.526.4107P}. However, many clusters are actually duplicates. This prevents us from using the database directly to search for binary clusters. It is necessary to carry out further studies after the database has been processed.


\section*{Data availability}
The data can be derived from Zenodo (DOI: 10.5281/zenodo.12703204).

\begin{acknowledgements}
This work has been supported by the National Natural Science Foundation of China (No. 12473029) Dali Expert Workstation of Rainer Spurzem, Yunnan Academician Workstation of Wang Jingxiu (202005AF150025), China Manned Space Project (No. CMS-CSST-2021-A08), and Guanghe Fundation (No. ghfund202407013470).
This work has made use of data from the European Space Agency (ESA) mission
{\it Gaia} (\url{https://www.cosmos.esa.int/gaia}), processed by the {\it Gaia}
Data Processing and Analysis Consortium (DPAC,
\url{https://www.cosmos.esa.int/web/gaia/dpac/consortium}). Funding for the DPAC
has been provided by national institutions, in particular the institutions
participating in the {\it Gaia} Multilateral Agreement.
\end{acknowledgements}

\onecolumn

\begin{appendix} \label{sec:appendix}
\section{Comparison of cluster parameters with recent literatures}

In the appendix, we compare the stellar ages and masses that are derived from isochrone fit in this work and the results of some recent literatures \citep{2024A&A...686A..42H,2023MNRAS.525.2315A,2024AJ....167...12C}.
This work estimates the total cluster mass including some stars that were not observed, e.g. very faint stars, and the initial mass is given, for each cluster. Meanwhile, most other works counted the current masses of the observed stars. In addition, different methods including isochrone fits, Monte Carlo method, and artificial neural network were used by these works. We see that the stellar ages from different works are usually different, and the masses from this work are significantly larger than other works for most clusters. However, it is reasonable because there should be much more low mass stars if the members of a cluster obey a exponential-like IMF.

\begin{table}[h!]
    \centering
    \caption{Cluster parameters from this work and some recent literatures.  }

    \begin{tabular}{cccccccccccccccc}
\hline\hline
\multicolumn{1}{c}{Cluster}	&	
\multicolumn{2}{c}{This work} &	
\multicolumn{3}{c}{HR23/24} &	
\multicolumn{1}{c}{AMD23} &	
\multicolumn{1}{c}{TSC22} &	
\multicolumn{1}{c}{CSC24}\\	
\cmidrule(lr){2-3} \cmidrule(lr){4-6} \cmidrule(lr){7-7} \cmidrule(lr){8-8} \cmidrule(lr){9-9}
	&	Age [Myr]	&	Mass [$M_\odot$]	&	Age [Myr]	&	Mass [$M_\odot$]	&	$r_t$ [pc]	&	Mass [$M_\odot$]	&	$r_t$ [pc]	&	Age [Myr] \cr
\hline
ASCC100	&	31	&	1664.59$\pm$118.79	&	65 	&	43.18$\pm$15.52	&	18.35$\pm$0.44	&		&		&	16 	\\
HSC534	&	31	&	352.19$\pm$9.22	&	220 	&	21.46$\pm$3.86	&	10.39$\pm$0.09	&		&		&	23 	\\
BDSB93	&	37	&	416.29$\pm$10.45	&	7 	&	168.58$\pm$23.38	&	17.24$\pm$0.14	&		&		&	81 	\\
HSC1740	&	13	&	18027.73$\pm$2279.65	&	8 	&	132.31$\pm$24.14	&	7.64$\pm$0.32	&		&		&	309 	\\
CWNU1024	&	21	&	7425.00$\pm$988.68	&	32 	&	236.76$\pm$33.45	&	61.10$\pm$2.71	&		&		&	11 	\\
OCSN82	&	26	&	1939.08$\pm$134.03	&	24 	&	143.9$\pm$15.91	&	29.6$\pm$0.68	&		&		&	13 	\\
CWNU1173	&	98	&	1864.56$\pm$266.66	&	100 	&	191.65$\pm$21.51	&	34.80$\pm$1.66	&		&		&	23 	\\
CWNU1256	&	98	&	928.03$\pm$268.96	&	74 	&	42.5$\pm$11.64	&	13.88$\pm$1.34	&		&		&	288 	\\
CWNU2666	&	1480	&	713.96$\pm$104.22	&	98 	&	447.85$\pm$59.99	&	6.99$\pm$0.34	&		&		&	54 	\\
HSC224	&	1410	&	816.00$\pm$461.53	&	64 	&	427.51$\pm$40.56	&	21.35$\pm$4.03	&		&		&	50 	\\
HSC189	&	98	&	621.52$\pm$50.97	&	125 	&	76.54$\pm$20.05	&	24.62$\pm$0.67	&		&		&	257	\\
UPK24	&	676	&	1040.68$\pm$138.86	&	471 	&	157.94$\pm$23.72	&	40.23$\pm$1.79	&	105$\pm$21	&	2.50$\pm$79.12	&	380 	\\
HSC1897	&	45	&	1421.53$\pm$201.28	&	89 	&	97.03$\pm$13.83	&	24.50$\pm$1.16	&		&		&	23 	\\
HSC1911	&	45	&	1527.33$\pm$54.87	&	130 	&	103.77$\pm$17.69	&	36.40$\pm$0.44	&		&		&	178 	\\
HSC2468	&	13	&	4984.30$\pm$861.13	&	8 	&	118.9$\pm$13.72	&	23.88$\pm$1.38	&		&		&	36 	\\
HSC2505	&	20	&	4930.71$\pm$464.42	&	7 	&	52.98$\pm$10.84	&	5.48$\pm$0.17	&		&		&	53 	\\
HSC2571	&	162	&	550.45$\pm$73.75	&	356 	&	30.54$\pm$9.36	&	14.33$\pm$0.64	&		&		&	282 	\\
Platais12	&	162	&	4059.79$\pm$116.79	&	147 	&	431.96$\pm$28.80	&	46.17$\pm$0.44	&		&		&	87 	\\
\hline
    \end{tabular}
    \tablefoot{HR24, AMD23, TSC22 and CSC24 refer to the catalogues of \cite{2024A&A...686A..42H}, \cite{2023MNRAS.525.2315A}, \cite{2022A&A...659A..59T}, and \cite{2024AJ....167...12C}.}
    \label{appendix.table}

\end{table}

\end{appendix}


\begin{thebibliography}{}

\bibitem[Allen et al.(2018)]{2018MNRAS.481.3953A} Allen, C., Ruelas-Mayorga, A., S{\'a}nchez, L.~J., et al.\ 2018, \mnras, 481, 3953

\bibitem[Almeida et al.(2023)]{2023MNRAS.525.2315A} Almeida, A., Monteiro, H., \& Dias, W.~S.\ 2023, \mnras, 525, 2315

\bibitem[{{Anders} {et~al.}(2022){Anders}, {Castro-Ginard}, {Casado}, {Jordi},
  \& {Balaguer-N{\'u}{\~n}ez}}]{2022RNAAS...6...58A}
{Anders}, F., {Castro-Ginard}, A., {Casado}, J., {Jordi}, C., \&
  {Balaguer-N{\'u}{\~n}ez}, L. 2022, Research Notes of the American
  Astronomical Society, 6, 58


\bibitem[{{Bekki} {et~al.}(2004){Bekki}, {Beasley}, {Forbes}, \&
  {Couch}}]{2004ApJ...602..730B}
{Bekki}, K., {Beasley}, M.~A., {Forbes}, D.~A., \& {Couch}, W.~J. 2004, \apj,
  602, 730

\bibitem[Bhatia \& Hatzidimitriou(1988)]{1988MNRAS.230..215B} Bhatia, R.~K. \& Hatzidimitriou, D.\ 1988, \mnras, 230, 215


\bibitem[{{Binney} \& {Tremaine}(2008)}]{2008gady.book.....B}
{Binney}, J., \& {Tremaine}, S. 2008, {Galactic Dynamics: Second Edition}


\bibitem[Bressan et al.(2012)]{2012MNRAS.427..127B} {Bressan}, A., {Marigo}, P., {Girardi}, L., {et~al.} 2012, \mnras, 427, 127


\bibitem[{{Brown} {et~al.}(1995){Brown}, {Burkert}, \&
  {Truran}}]{1995ApJ...440..666B}
{Brown}, J.~H., {Burkert}, A., \& {Truran}, J.~W. 1995, \apj, 440, 666


\bibitem[{{Camargo}(2021)}]{2021ApJ...923...21C}
{Camargo}, D. 2021, \apj, 923, 21

\bibitem[{{Camargo} {et~al.}(2016){Camargo}, {Bica}, \&
  {Bonatto}}]{2016MNRAS.455.3126C}
{Camargo}, D., {Bica}, E., \& {Bonatto}, C. 2016, \mnras, 455, 3126

\bibitem[Cantat-Gaudin \& Anders(2020)]{2020A&A...633A..99C} Cantat-Gaudin, T. \& Anders, F.\ 2020, \aap, 633, A99

\bibitem[{{Cantat-Gaudin} {et~al.}(2020){Cantat-Gaudin}, {Anders},
  {Castro-Ginard}, {Jordi}, {Romero-G{\'o}mez}, {Soubiran}, {Casamiquela},
  {Tarricq}, {Moitinho}, {Vallenari}, {Bragaglia}, {Krone-Martins}, \&
  {Kounkel}}]{2020AA...640A...1C}
{Cantat-Gaudin}, T., {Anders}, F., {Castro-Ginard}, A., {et~al.} 2020, \aap,
  640, A1

\bibitem[Casagrande \& VandenBerg(2018)]{2018MNRAS.479L.102C} Casagrande, L. \& VandenBerg, D.~A.\ 2018, \mnras, 479, L102

\bibitem[Cavallo et al.(2024)]{2024AJ....167...12C} Cavallo, L., Spina, L., Carraro, G., et al.\ 2024, \aj, 167, 12

\bibitem[{{Conrad} {et~al.}(2017){Conrad}, {Scholz}, {Kharchenko}, {Piskunov},
  {R{\"o}ser}, {Schilbach}, {de Jong}, {Schnurr}, {Steinmetz}, {Grebel},
  {Zwitter}, {Bienaym{\'e}}, {Bland-Hawthorn}, {Gibson}, {Gilmore},
  {Kordopatis}, {Kunder}, {Navarro}, {Parker}, {Reid}, {Seabroke}, {Siviero},
  {Watson}, \& {Wyse}}]{2017AA...600A.106C}
{Conrad}, C., {Scholz}, R.~D., {Kharchenko}, N.~V., {et~al.} 2017, \aap, 600,
  A106


\bibitem[{{de La Fuente Marcos} \& {de La Fuente
  Marcos}(2009)}]{2009AA...500L..13D}
{de La Fuente Marcos}, R., \& {de La Fuente Marcos}, C. 2009, \aap, 500, L13


\bibitem[de la Fuente Marcos \& de la Fuente Marcos(2010)]{2010ApJ...719..104D} de la Fuente Marcos, R. \& de la Fuente Marcos, C.\ 2010, \apj, 719, 104

\bibitem[Deng \& Li(2024)]{2024RAA....24f5004D} Deng, Y.-Y. \& Li, Z.-M.\ 2024, Research in Astronomy and Astrophysics, 24, 065004

\bibitem[De Silva et al.(2015)]{2015MNRAS.453..106D} De Silva, G.~M., Carraro, G., D'Orazi, V., et al.\ 2015, \mnras, 453, 106

\bibitem[{{Dias} {et~al.}(2002){Dias}, {Alessi}, {Moitinho}, \&
  {L{\'e}pine}}]{2002AA...389..871D}
{Dias}, W.~S., {Alessi}, B.~S., {Moitinho}, A., \& {L{\'e}pine}, J.~R.~D. 2002,
  \aap, 389, 871

\bibitem[{{Dias} {et~al.}(2021){Dias}, {Monteiro}, {Moitinho}, {L{\'e}pine},
  {Carraro}, {Paunzen}, {Alessi}, \& {Villela}}]{2021MNRAS.504..356D}
{Dias}, W.~S., {Monteiro}, H., {Moitinho}, A., {et~al.} 2021, \mnras, 504, 356

\bibitem[{{Dieball} {et~al.}(2002){Dieball}, {M{\"u}ller}, \&
  {Grebel}}]{2002AA...391..547D}
{Dieball}, A., {M{\"u}ller}, H., \& {Grebel}, E.~K. 2002, \aap, 391, 547


\bibitem[{{Fujimoto} \& {Kumai}(1997)}]{1997AJ....113..249F}
{Fujimoto}, M., \& {Kumai}, Y. 1997, \aj, 113, 249

\bibitem[Gaia Collaboration et al.(2021)]{2021A&A...649A...1G} Gaia Collaboration, Brown, A.~G.~A., Vallenari, A., et al.\ 2021, \aap, 649, A1

\bibitem[{{Gaia Collaboration} {et~al.}(2023){Gaia Collaboration}, {Drimmel},
  {Romero-G{\'o}mez}, {Chemin}, {Ramos}, {Poggio}, {Ripepi}, {Andrae},
  {Blomme}, {Cantat-Gaudin}, {Castro-Ginard}, {Clementini}, {Figueras},
  {Fouesneau}, {Fr{\'e}mat}, {Jardine}, {Khanna}, {Lobel}, {Marshall},
  {Muraveva}, {Brown}, {Vallenari}, {Prusti}, {de Bruijne}, {Arenou},
  {Babusiaux}, {Biermann}, {Creevey}, {Ducourant}, {Evans}, {Eyer}, {Guerra},
  {Hutton}, {Jordi}, {Klioner}, {Lammers}, {Lindegren}, {Luri}, {Mignard},
  {Panem}, {Pourbaix}, {Randich}, {Sartoretti}, {Soubiran}, {Tanga}, {Walton},
  {Bailer-Jones}, {Bastian}, {Jansen}, {Katz}, {Lattanzi}, {van Leeuwen},
  {Bakker}, {Cacciari}, {Casta{\~n}eda}, {De Angeli}, {Fabricius}, {Galluccio},
  {Guerrier}, {Heiter}, {Masana}, {Messineo}, {Mowlavi}, {Nicolas},
  {Nienartowicz}, {Pailler}, {Panuzzo}, {Riclet}, {Roux}, {Seabroke}, {Sordo},
  {Th{\'e}venin}, {Gracia-Abril}, {Portell}, {Teyssier}, {Altmann}, {Audard},
  {Bellas-Velidis}, {Benson}, {Berthier}, {Burgess}, {Busonero}, {Busso},
  {C{\'a}novas}, {Carry}, {Cellino}, {Cheek}, {Damerdji}, {Davidson}, {de
  Teodoro}, {Nu{\~n}ez Campos}, {Delchambre}, {Dell'Oro}, {Esquej},
  {Fern{\'a}ndez-Hern{\'a}ndez}, {Fraile}, {Garabato}, {Garc{\'\i}a-Lario},
  {Gosset}, {Haigron}, {Halbwachs}, {Hambly}, {Harrison}, {Hern{\'a}ndez},
  {Hestroffer}, {Hodgkin}, {Holl}, {Jan{\ss}en}, {Jevardat de Fombelle},
  {Jordan}, {Krone-Martins}, {Lanzafame}, {L{\"o}ffler}, {Marchal}, {Marrese},
  {Moitinho}, {Muinonen}, {Osborne}, {Pancino}, {Pauwels}, {Recio-Blanco},
  {Reyl{\'e}}, {Riello}, {Rimoldini}, {Roegiers}, {Rybizki}, {Sarro}, {Siopis},
  {Smith}, {Sozzetti}, {Utrilla}, {van Leeuwen}, {Abbas}, {{\'A}brah{\'a}m},
  {Abreu Aramburu}, {Aerts}, {Aguado}, {Ajaj}, {Aldea-Montero}, {Altavilla},
  {{\'A}lvarez}, {Alves}, {Anders}, {Anderson}, {Anglada Varela}, {Antoja},
  {Baines}, {Baker}, {Balaguer-N{\'u}{\~n}ez}, {Balbinot}, {Balog}, {Barache},
  {Barbato}, {Barros}, {Barstow}, {Bartolom{\'e}}, {Bassilana}, {Bauchet},
  {Becciani}, {Bellazzini}, {Berihuete}, {Bernet}, {Bertone}, {Bianchi},
  {Binnenfeld}, {Blanco-Cuaresma}, {Boch}, {Bombrun}, {Bossini}, {Bouquillon},
  {Bragaglia}, {Bramante}, {Breedt}, {Bressan}, {Brouillet}, {Brugaletta},
  {Bucciarelli}, {Burlacu}, {Butkevich}, {Buzzi}, {Caffau}, {Cancelliere},
  {Carballo}, {Carlucci}, {Carnerero}, {Carrasco}, {Casamiquela}, {Castellani},
  {Chaoul}, {Charlot}, {Chiaramida}, {Chiavassa}, {Chornay}, {Comoretto},
  {Contursi}, {Cooper}, {Cornez}, {Cowell}, {Crifo}, {Cropper}, {Crosta},
  {Crowley}, {Dafonte}, {Dapergolas}, {David}, {de Laverny}, {De Luise}, {De
  March}, {De Ridder}, {de Souza}, {de Torres}, {del Peloso}, {del Pozo},
  {Delbo}, {Delgado}, {Delisle}, {Demouchy}, {Dharmawardena}, {Di Matteo},
  {Diakite}, {Diener}, {Distefano}, {Dolding}, {Enke}, {Fabre}, {Fabrizio},
  {Faigler}, {Fedorets}, {Fernique}, {Fournier}, {Fouron}, {Fragkoudi}, {Gai},
  {Garcia-Gutierrez}, {Garcia-Reinaldos}, {Garc{\'\i}a-Torres}, {Garofalo},
  {Gavel}, {Gavras}, {Gerlach}, {Geyer}, {Giacobbe}, {Gilmore}, {Girona},
  {Giuffrida}, {Gomel}, {Gomez}, {Gonz{\'a}lez-N{\'u}{\~n}ez},
  {Gonz{\'a}lez-Santamar{\'\i}a}, {Gonz{\'a}lez-Vidal}, {Granvik}, {Guillout},
  {Guiraud}, {Guti{\'e}rrez-S{\'a}nchez}, {Guy}, {Hatzidimitriou}, {Hauser},
  {Haywood}, {Helmer}, {Helmi}, {Sarmiento}, {Hidalgo}, {H{\l}adczuk}, {Hobbs},
  {Holland}, {Huckle}, {Jasniewicz}, {Jean-Antoine Piccolo},
  {Jim{\'e}nez-Arranz}, {Juaristi Campillo}, {Julbe}, {Karbevska}, {Kervella},
  {Kordopatis}, {Korn}, {K{\'o}sp{\'a}l}, {Kostrzewa-Rutkowska},
  {Kruszy{\'n}ska}, {Kun}, {Laizeau}, {Lambert}, {Lanza}, {Lasne}, {Le
  Campion}, {Lebreton}, {Lebzelter}, {Leccia}, {Leclerc}, {Lecoeur-Taibi},
  {Liao}, {Licata}, {Lindstr{\o}m}, {Lister}, {Livanou}, {Lorca}, {Loup},
  {Madrero Pardo}, {Magdaleno Romeo}, {Managau}, {Mann}, {Manteiga},
  {Marchant}, {Marconi}, {Marcos}, {Marcos Santos}, {Mar{\'\i}n Pina},
  {Marinoni}, {Marocco}, {Martin Polo}, {Mart{\'\i}n-Fleitas}, {Marton},
  {Mary}, {Masip}, {Massari}, {Mastrobuono-Battisti}, {Mazeh}, {McMillan},
  {Messina}, {Michalik}, {Millar}, {Mints}, {Molina}, {Molinaro}, {Moln{\'a}r},
  {Monari}, {Mongui{\'o}}, {Montegriffo}, {Montero}, {Mor}, {Mora},
  {Morbidelli}, {Morel}, {Morris}, {Murphy}, {Musella}, {Nagy}, {Noval},
  {Oca{\~n}a}, {Ogden}, {Ordenovic}, {Osinde}, {Pagani}, {Pagano}, {Palaversa},
  {Palicio}, {Pallas-Quintela}, {Panahi}, {Payne-Wardenaar}, {Pe{\~n}alosa
  Esteller}, {Penttil{\"a}}, {Pichon}, {Piersimoni}, {Pineau}, {Plachy},
  {Plum}, {Pr{\v{s}}a}, {Pulone}, {Racero}, {Ragaini}, {Rainer}, {Raiteri},
  {Ramos-Lerate}, {Re Fiorentin}, {Regibo}, {Richards}, {Rios Diaz}, {Riva},
  {Rix}, {Rixon}, {Robichon}, {Robin}, {Robin}, {Roelens}, {Rogues},
  {Rohrbasser}, {Rowell}, {Royer}, {Ruz Mieres}, {Rybicki}, {Sadowski},
  {S{\'a}ez N{\'u}{\~n}ez}, {Sagrist{\`a} Sell{\'e}s}, {Sahlmann}, {Salguero},
  {Samaras}, {Sanchez Gimenez}, {Sanna}, {Santove{\~n}a}, {Sarasso},
  {Schultheis}, {Sciacca}, {Segol}, {Segovia}, {S{\'e}gransan}, {Semeux},
  {Shahaf}, {Siddiqui}, {Siebert}, {Siltala}, {Silvelo}, {Slezak}, {Slezak},
  {Smart}, {Snaith}, {Solano}, {Solitro}, {Souami}, {Souchay}, {Spagna},
  {Spina}, {Spoto}, {Steele}, {Steidelm{\"u}ller}, {Stephenson}, {S{\"u}veges},
  {Surdej}, {Szabados}, {Szegedi-Elek}, {Taris}, {Taylor}, {Teixeira},
  {Tolomei}, {Tonello}, {Torra}, {Torra}, {Torralba Elipe}, {Trabucchi},
  {Tsounis}, {Turon}, {Ulla}, {Unger}, {Vaillant}, {van Dillen}, {van Reeven},
  {Vanel}, {Vecchiato}, {Viala}, {Vicente}, {Voutsinas}, {Weiler}, {Wevers},
  {Wyrzykowski}, {Yoldas}, {Yvard}, {Zhao}, {Zorec}, {Zucker}, \&
  {Zwitter}}]{2023AA...674A..37G}
{Gaia Collaboration}, {Drimmel}, R., {Romero-G{\'o}mez}, M., {et~al.} 2023,
  \aap, 674, A37

\bibitem[Gaia Collaboration et al.(2023)]{2023A&A...674A...1G} Gaia Collaboration, Vallenari, A., Brown, A.~G.~A., et al.\ 2023, \aap, 674, A1

\bibitem[{{Hatzidimitriou} \& {Bhatia}(1990)}]{1990AA...230...11H}
{Hatzidimitriou}, D., \& {Bhatia}, R.~K. 1990, \aap, 230, 11

\bibitem[{{He} {et~al.}(2022{\natexlab{a}}){He}, {Wang}, {Luo}, {Li}, {Liu}, \&
  {Jiang}}]{2022ApJS..262....7H}
{He}, Z., {Wang}, K., {Luo}, Y., {et~al.} 2022{\natexlab{a}}, \apjs, 262, 7

\bibitem[He et al.(2023)]{2023ApJS..264....8H} He, Z., Liu, X., Luo, Y., et al.\ 2023, \apjs, 264, 8

\bibitem[{{Hunt} \& {Reffert}(2023)}]{2023AA...673A.114H}
{Hunt}, E.~L., \& {Reffert}, S. 2023, \aap, 673, A114

\bibitem[Hunt \& Reffert(2024)]{2024A&A...686A..42H} Hunt, E.~L. \& Reffert, S.\ 2024, \aap, 686, A42

\bibitem[King(1962)]{1962AJ.....67..471K}
{King}, I.\ 1962, \aj, 67, 471

\bibitem[Kroupa(2001)]{2001MNRAS.322..231K} Kroupa, P.\ 2001, \mnras, 322, 231

\bibitem[Li \& Mao(2024)]{2024RAA....24e5014L} Li, Z.-M. \& Mao, C.-Y.\ 2024, Research in Astronomy and Astrophysics, 24, 055014

\bibitem[{{Li} {et~al.}(2017){Li}, {Mao}, {Luo}, {Fan}, {Zhao}, {Chen}, {Li},
  \& {Guo}}]{2017RAA....17...71L}
{Li}, Z.-M., {Mao}, C.-Y., {Luo}, Q.-P., {et~al.} 2017, Research in Astronomy
  and Astrophysics, 17, 071

\bibitem[Kuhn et al.(2019)]{2019ApJ...870...32K} Kuhn, M.~A., Hillenbrand, L.~A., Sills, A., et al.\ 2019, \apj, 870, 32

\bibitem[Loktin(1997)]{1997A&AT...14..181L} Loktin, A.~V.\ 1997, Astronomical and Astrophysical Transactions, 14, 181

\bibitem[{{Loktin} \& {Popova}(2017)}]{2017AstBu..72..257L}
{Loktin}, A.~V., \& {Popova}, M.~E. 2017, Astrophysical Bulletin, 72, 257

\bibitem[{{Lynga}(1995)}]{1995yCat.7092....0L}
{Lynga}, G. 1995, VizieR Online Data Catalogue, VII/92A

\bibitem[McNamara \& Sanders(1983)]{1983A&A...118..361M}
{McNamara}, B.~J. \& {Sanders}, W.~L.\ 1983, \aap, 118, 361

\bibitem[{{Mermilliod} \& {Paunzen}(2003)}]{2003AA...410..511M}
{Mermilliod}, J.~C., \& {Paunzen}, E. 2003, \aap, 410, 511

\bibitem[Minniti et al.(2004)]{2004ApJ...612..215M} Minniti, D., Rejkuba, M., Funes, J.~G., et al.\ 2004, \apj, 612, 215

\bibitem[Perren et al.(2023)]{2023MNRAS.526.4107P} Perren, G.~I., Pera, M.~S., Navone, H.~D., et al.\ 2023, \mnras, 526, 4107

\bibitem[{{Piecka} \& {Paunzen}(2021)}]{2021AA...649A..54P}
{Piecka}, M., \& {Paunzen}, E. 2021, \aap, 649, A54

\bibitem[Piskunov et al.(2008)]{2008A&A...487..557P}
{Piskunov}, A.~E., {Kharchenko}, N.~V., {Schilbach}, E., et al.\ 2008, \aap, 487, 557

\bibitem[Portegies Zwart \& Rusli(2007)]{2007MNRAS.374..931P} Portegies Zwart, S.~F. \& Rusli, S.~P.\ 2007, \mnras, 374, 931

\bibitem[Portegies Zwart et al.(2010)]{2010ARA&A..48..431P} Portegies Zwart, S.~F., McMillan, S.~L.~W., \& Gieles, M.\ 2010, \araa, 48, 431

\bibitem[Priyatikanto et al.(2016)]{2016MNRAS.457.1339P} Priyatikanto, R., Kouwenhoven, M.~B.~N., Arifyanto, M.~I., et al.\ 2016, \mnras, 457, 1339

\bibitem[Qin et al.(2023)]{2023ApJS..265...12Q} Qin, S., Zhong, J., Tang, T., et al.\ 2023, \apjs, 265, 12

\bibitem[Skrutskie et al.(2006)]{2006AJ...131..1163}
Skrutskie, M. F., Cutri, R. M., Stiening, R. C., et al. (2006). The Two Micron All-Sky Survey (2MASS): A Survey of the Universe's Oldest Light. Astronomical Journal, 131(2), 1163-1183

\bibitem[{{Song} {et~al.}(2022){Song}, {Esamdin}, {Hu}, \&
  {Zhang}}]{2022AA...666A..75S}
{Song}, F., {Esamdin}, A., {Hu}, Q., \& {Zhang}, M. 2022, \aap, 666, A75

\bibitem[{{Subramaniam} {et~al.}(1995){Subramaniam}, {Gorti}, {Sagar}, \&
  {Bhatt}}]{1995AA...302...86S}
{Subramaniam}, A., {Gorti}, U., {Sagar}, R., \& {Bhatt}, H.~C. 1995, \aap, 302, 86

\bibitem[Tarricq et al.(2022)]{2022A&A...659A..59T} Tarricq, Y., Soubiran, C., Casamiquela, L., et al.\ 2022, \aap, 659, A59

\bibitem[{{Weisz} {et~al.}(2015)}]{2015ApJ...806..198W}
{Weisz}, Daniel R., {Johnson}, L. Clifton, {Foreman-Mackey}, Daniel, {Dolphin}, Andrew E., et~al. \apj, 115, 2384

\end{thebibliography}
\end{document}